\documentclass[twocolumn,secnumarabic,amssymb, nofootinbib, nobibnotes, aip, prd, reprint,superscriptaddress]{revtex4-1}

\usepackage{graphics}      
\usepackage{bm}            
\usepackage[cmex10]{amsmath}
\usepackage{amssymb}
\usepackage{mathrsfs}
\usepackage{mathtools}
\usepackage[british]{babel}
\usepackage[titletoc,toc,title]{appendix}
\usepackage{scrextend}

\usepackage{color}

\begin{document}
\title{Laser-plasma interaction in magnetized environment}
\author         {Yuan Shi}
\email          {yshi@pppl.gov}
\affiliation    {Department of Astrophysical Sciences, Princeton University, Princeton, NJ 08544 USA}
\affiliation    {Princeton Plasma Physics Laboratory, Princeton University, Princeton, NJ 08543 USA}

\author         {Hong Qin}
\affiliation    {Department of Astrophysical Sciences, Princeton University, Princeton, NJ 08544 USA}
\affiliation    {Princeton Plasma Physics Laboratory, Princeton University, Princeton, NJ 08543 USA}
\affiliation    {School of Nuclear Science and Technology and Department of Modern Physics, University of Science and Technology of China, Hefei, Anhui 230026, China}

\author         {Nathaniel J. Fisch}
\affiliation    {Department of Astrophysical Sciences, Princeton University, Princeton, NJ 08544 USA}
\affiliation    {Princeton Plasma Physics Laboratory, Princeton University, Princeton, NJ 08543 USA}

\date           {\today}

\begin{abstract}
Propagation and scattering of lasers present new phenomena and applications when the plasma medium becomes strongly magnetized. %
With mega-Gauss magnetic fields, scattering of optical lasers already becomes manifestly anisotropic. Special angles exist where coherent laser scattering is either enhanced or suppressed, as we demonstrate using a cold-fluid model. Consequently, by aiming laser beams at special angles, one may be able to optimize laser-plasma coupling in magnetized implosion experiments. %
In addition, magnetized scattering can be exploited to improve the performance of plasma-based laser pulse amplifiers. Using the magnetic field as an extra control variable, it is possible to produce optical pulses of higher intensity, as well as compress UV and soft x-ray pulses beyond the reach of other methods. %
In even stronger giga-Gauss magnetic fields, laser-plasma interactions begin to enter the relativistic-quantum regime. Using quantum electrodynamics, we compute modified wave dispersion relation, which enables correct interpretation of Faraday rotation measurements of strong magnetic fields. 
\end{abstract}

\maketitle
\setlength{\parskip}{0pt}
\section{Introduction}
Magnetic fields affect laser propagation and scattering when the electron gyrofrequency $\Omega_e=eB/m_e$ is no longer ignorable compared to the laser frequency. For example, a magnetic field $\sim10$ MG, corresponding to $\Omega_e\hbar\sim0.1$ eV, will noticeably alter the wave dispersion relation and the scattering cross section of optical lasers in plasmas. In low density plasmas, the role of the strong magnetic field is largely classical. However, as plasma density increases, quantum effects may emerge when the characteristic size of electron wave functions $l_B=\sqrt{2\hbar/eB}$ becomes comparable to inter-particle spacing. For example, a magnetic field $\sim10$ MG, corresponding to the magnetic de Broglie wavelength $l_B\sim 1$ nm, may already allow electrons to feel the Fermi degeneracy in solid-density plasmas. As the field strength further increases towards the Schwinger limit $B\sim10^{13}$ G, where the magnetic de Broglie wavelength shrinks to electron Compton wave length, relativistic-quantum effects of magnetic fields will become increasingly prominent. 

While magnetic fields on the order of Schwinger limit can only be found near neutron stars, mega-Gauss to giga-Gauss magnetic fields can already be produced by a number of laboratory techniques. For example, using lasers to drive plasma implosions, seed magnetic fields, either self-generated \cite{Igumenshchev14} or externally imposed \cite{Gotchev09,Knauer10}, can be amplified to tens of mega-Gauss by magnetic flux compression. A more controllable technique produces magnetic fields of similar strengths using lasers to drive capacitor-coil targets \cite{Fujioka13,Santos15,Goyon17}. Comparable or even stronger magnetic fields can be produced by dynamo effects when solid targets are directly ablated by intense laser pulses \cite{Borghesi98,Tatarakis02,Tatarakis02Nature,Wagner04,Manuel12,Gao12,Gao15}. Using these techniques, magnetic fields may be further intensified by employing stronger drive lasers. This emerging availability of very strong magnetic fields thus present new challenges and opportunities that remain to be investigated. In this paper, we review three research directions where effects of strong magnetic fields during laser-plasma interactions have been explored.

The first direction is coherent scattering of lasers in plasmas, where magnetic fields on the order of mega-Gauss can make noticeable differences for 1 $\mu$m lasers. Coherent scattering happens when the Debye length of the plasma is not much larger than the laser wavelength. In this case, instead of interacting directly with individual charged particles \cite{Sheffield10}, lasers interact collectively with plasma waves and scatter due to nonlinear motion of the plasma. Magnetized waves that scatter lasers are Alfv\'en waves, hybrid waves, and Bernstein waves \cite{Stix92}, which replace the Langmuir wave and the ion-acoustic wave in unmagnetized plasmas. Consequently, Raman and Brillouin scattering \cite{Drake74}, the two coherent scattering modes in unmagnetized plasmas, are now replaced by scattering mediated by magnetized plasma waves, on which the magnetic anisotropy is imprinted  \cite{Shi17scatter}. Understanding angular dependences due to the anisotropy is especially important for magnetized laser implosion experiments \cite{Wang15,Barnak17}, where multiple laser beams propagate at angles with respect to the magnetic field.    

The second direction is laser pulse compression mediated by magnetized plasmas, for which tens of mega-Gauss magnetic fields start to bring significant improvements when amplifying 1 $\mu$m lasers. During laser pulse compression \cite{Milroy79}, energy stored in a long pump laser is transferred, via a plasma wave, to a seed pulse, whose intensity is amplified and duration is shortened. While using Raman compression \cite{Malkin99,Malkin07} and Brillouin compression \cite{Andreev06,Edwards16,Edwards17} in unmagnetized plasmas, intense laser pulses beyond the reach of Chirped Pulse Amplification \cite{Maine88} may already be produced, magnetic fields bring additional improvements\cite{Shi17laser}. For optical lasers, the improvements are primarily engineering, where external magnetic fields allow better control of the plasma uniformity. On the other hand, for shorter wavelength lasers, the improvements due to alleviation of physical constraints, such as damping and instabilities, become more substantial. Due to these improvements, magnetized mediations may be used to compress intense UV and soft x-ray pulses, which cannot be compressed using other methods.   

The third direction is laser propagation, which remains largely classical until magnetic fields on the order of giga-Gauss are present. Although giga-Gauss magnetic fields are still far below the Schwinger limit, relativistic quantum effects may already be observable when they are boosted near singularities. For example, relativistic quantum effects can noticeably alter the dependency of Faraday rotation on the frequency of lasers, especially when the frequency approaches the cutoff frequency of the right-circularly-polarized (\textit{R}) wave. When approaching the \textit{R} wave cutoff, the phase velocity of \textit{R} wave goes to infinity, while the phase velocity of the left-circularly-polarized (\textit{L}) wave remains finite. Therefore, the difference in these phase velocities, which leads to Faraday rotation of linearly polarized lasers, becomes singular. This singularity boosts relativistic quantum effects and can produces order unity corrections to Faraday rotation in strongly magnetized plasmas \cite{Shi16QED}. Not surprisingly, when the magnetic field becomes even stronger, relativistic-quantum effects will become more appreciable. 

This paper reviews progress made in these three research directions and motivates future endeavors towards understanding and utilizing magnetic fields during laser-plasma interactions. In Sec.~\ref{sec:scatter}, we present coherent laser scattering phenomena in magnetized plasmas, using results from a cold fluid theory as illustrations. In Sec.~\ref{sec:amplification}, we demonstrate applications of strong magnetic fields, using laser pulse compression mediated by the upper-hybrid wave as an example. In Sec.~\ref{sec:propagation}, we discuss the new physical regime that strong magnetic fields enable us to reach, by studying how relativistic-quantum effects modify Faraday rotation 
as an example. In Sec.~\ref{sec:conclusion}, we summarize challenges and opportunities, as strong magnetic fields become available.

\section{\label{sec:scatter}Coherent three-wave scattering}
Coherent scattering is a primary way by which long wavelength lasers are scattered in high-density plasmas. To put this type of scattering in the context of other scattering mechanisms, note that the degree of coherence of laser scattering can vary, depending on the wavelength of the laser. Incoherent scattering is the extreme where the wavelength of the laser is much smaller than the electron correlation length. In this case, the laser resolves the discreteness of the medium and directly wiggles individual particles, which radiate secondary electromagnetic waves as scattered lights. In the other extreme, coherent scattering happens when the wavelength of the laser is much larger than the electron correlation length. In this case, motion of charged particles is highly synchronized, and the laser scatters due to collective nonlinear response of the plasma medium. In this section, we will focus on coherent scattering of an incident laser due to resonant three-wave interactions.  

\subsection{Resonant Three-wave Interactions}
The lowest order nonlinearities couple three waves, and resonant three-wave interactions can happen when frequencies $\omega_i$ and wave vectors $\mathbf{k}_i$ of the three waves satisfy the resonant conditions
\begin{eqnarray}
\label{eq:resonanceW}
\omega_1&=&\omega_2+\omega_3, \\
\label{eq:resonanceK}
\mathbf{k}_1&=&\mathbf{k}_2+\mathbf{k}_3,
\end{eqnarray}
where all $\omega_i$'s are positive. These resonant conditions only need to be satisfied approximately, because large amplitude waves can have finite band width. Moreover, when two of these three waves are strongly driven by external sources, the third wave does not need to be a linear eigenmode of the system.

When all three waves are eigenmodes of the homogeneous system, their envelopes evolve slowly due to their weak resonant coupling. The evolution of wave envelopes can be described by the three-wave equations \cite{Kaup79}
\begin{eqnarray}
\label{eq:3waves1}
d_{t}a_1&=&-\frac{\Gamma}{\omega_1}a_2a_3,\\
\label{eq:3waves2}
d_{t}a_2&=&\phantom{+}\frac{\Gamma}{\omega_2}a_3a_1,\\
\label{eq:3waves3}
d_{t}a_3&=&\phantom{+}\frac{\Gamma}{\omega_3}a_1a_2,
\end{eqnarray}
where $d_ta_i:=(\partial_t+\mathbf{v}_{gi}\cdot\nabla+\nu_i)a_i$ denotes the advective derivative. In the above equations, the real-valued $a_i=eE_iu_i^{1/2}/m_ec\omega_i$ is the normalized wave electric field, where $u_i$ is the coefficient such that the averaged energy of the linear wave is $U_i=\epsilon_0u_iE_i^2/2$. The normalized wave amplitude $a_i$ is advected at the wave group velocity $\mathbf{v}_{gi}=\partial\omega_i/\partial\mathbf{k}_i$, and is damped at a rate $\nu_i$. As the waves advect, they transfer energy between one another at a rate determined by $\Gamma$, the coupling coefficient. 


\subsection{Coupling Coefficient}
While resonant three-wave interactions can always be described by the same three-wave equations, the coupling coefficient is what encodes the physical details. This essential coefficient was very difficult to compute in the presence of a background magnetic field. Although many methods were attempted \cite{Sjolund67,Galloway71,Boyd78,Shivamoggi82,Ram82,Boyd85,Cohen87,Stefan87}, explicit expressions of the coupling coefficient were only known in the simple cases where the collimated waves propagate either parallel \cite{Laham98} or perpendicular \cite{Sanuki77,Grebogi80,Dodin17} to the magnetic field. Recently, we have obtained a convenient formula for the coupling coefficient in magnetized plasmas when waves propagate at arbitrary angles \cite{Shi17scatter}
\begin{equation}
\label{eq:coupling}
\Gamma=\sum_s\frac{Z_s\omega_{ps}^2\Theta^s_r}{4M_s(u_1u_2u_3)^{1/2}}.
\end{equation}
In the above formula, $Z_s:=e_s/e$ and $M_s:=m_s/m_e$ are the normalized charge and mass of species $s$, and $\omega_{ps}$ is its plasma frequency. 

The most important term in the coupling coefficient is $\Theta^s_r$, the real part of the normalized scattering strength 
\begin{eqnarray}
\label{eq:Theta3}
\nonumber
\Theta^s&=&\Theta_{1,\bar{2}\bar{3}}^s+\Theta_{\bar{2},\bar{3}1}^s+\Theta_{\bar{3},1\bar{2}}^s\\
&+&\Theta_{1,\bar{3}\bar{2}}^s+\Theta_{\bar{2},1\bar{3}}^s+\Theta_{\bar{3},\bar{2}1}^s.
\end{eqnarray}
This linear superposition of the strengths of six scattering channels corresponds to $3!=6$ ways the three waves can couple through the interaction Lagrangian. In Eq.~(\ref{eq:Theta3}), we used notations $\omega_{\bar{j}}=-\omega_j$, $\mathbf{k}_{\bar{j}}=-\mathbf{k}_j$, and $\mathbf{e}_{\bar{j}}=\mathbf{e}_j^*$. Using these notations, the normalized strength of each scattering channel is given by the simple expression
\begin{equation}
\label{eq:Thetaijl}
\Theta_{i,jl}^s=\frac{1}{\omega_j}(c\mathbf{k}_i\cdot\mathbb{F}^s_{j}\mathbf{e}_j)(\mathbf{e}_i\cdot\mathbb{F}^s_{l}\mathbf{e}_l),
\end{equation}
where $\mathbf{e}_j$ is the complex unit polarization vector of the $j$-th wave.  Notice that the relative phases of the three waves are important. 
The maximum coupling is attained when $\Theta_r=|\Theta|$. This happens when the phases of $\mathbf{e}_j$'s are synchronized to give the dominant scattering. 

The forcing operator $\mathbb{F}^s_{j}$, which appears in the scattering strengths, is related to the linear susceptibility $\chi^s_{j}$ by $\chi^s_{j}=-\omega_{ps}^2\mathbb{F}^s_{j}/\omega^2_{j}$. For example, in a cold-fluid plasma, the forcing operator is such that
\begin{equation}\label{eq:F}
\mathbb{F}^s_{j}\mathbf{z}=\gamma_{s,j}^2[\mathbf{z}+i\beta_{s,j}\mathbf{z}\times\mathbf{b} -\beta_{s,j}^2(\mathbf{z}\cdot\mathbf{b})\mathbf{b}],
\end{equation}
for any complex vector $\mathbf{z}\in\mathbb{C}^{3}$. Here, $\mathbf{b}$ is the unit vector along the background magnetic field, $\gamma_{s,j}^2:=1/(1-\beta_{s,j}^2)$ is the magnetization factor, and $\beta_{s,j}:=\Omega_{s}/\omega_{j}$ is the magnetization ratio. 
When thermal effects are of interest, one may replace the above cold-fluid forcing operator using the warm plasma susceptibility.

Finally, the last set of terms in 
the coupling coefficient are the wave energy coefficients
\begin{equation}
\label{eq:Ucoef}
u_j=\frac{1}{2}\mathbf{e}_j^\dagger\mathbb{H}_j\mathbf{e}_j.
\end{equation} 
Here, $\mathbb{H}_j=\partial(\omega_j^2\epsilon_j)/\omega_j\partial\omega_j$ is the wave energy operator, where $\epsilon_j=1+\sum_s\chi_j^s$ is the dielectric tensor. In terms of the forcing operator, the wave energy operator can be written as  
\begin{eqnarray}
\label{eq:Hk}
\mathbb{H}_{j}&=&2\mathbb{I}-\sum_s\frac{\omega_{ps}^2}{\omega_j} \frac{\partial\mathbb{F}^s_{j}}{\partial\omega_j}.
\end{eqnarray}
Using the above formulas, the coupling coefficient between any three resonant eigenmodes can be readily evaluated in the most general geometry.

\subsection{Experimental Observables}
To illustrate how the coupling coefficient can be related to experimental observables, let us consider Stokes scattering. As the incident laser $a_1$ propagates, it may pump the growth of some fluctuations $a_3$ in the plasma, while being scattered into $a_2$ as a frequency down-shifted laser. In the linear stage, the pump amplitude is roughly constant, and this Stokes scattering results in parametric growth of the scattered laser at an exponential rate
\begin{equation}
\label{eq:GrowthRate}
\gamma_0=\frac{|\Gamma a_1|}{\sqrt{\omega_2\omega_3}},
\end{equation} 
when damping and spatial variations are ignorable. To get a sense of how large this growth rate is, we can compare it with Raman scattering $\gamma_0=\gamma_R\mathcal{M}$, where the normalized growth rate
\begin{equation}
\mathcal{M}=2\frac{|\Gamma|}{\omega_p^2}\Big(\frac{\omega_p^3}{\omega_1\omega_2\omega_3}\Big)^{1/2},
\end{equation}
and $\gamma_R=\sqrt{\omega_1\omega_p}|a_1|/2$ the backward Raman growth rate in an unmagnetized plasma of the same density. Here, $\omega_p^2=\sum_s\omega_{ps}^2$ is the total plasma frequency. In experiments, dominant signals will come from largest growth rates, which emerge when wave phases are synchronized.

To evaluate the growth rate, we can imagine what happens in an experiment, in which the frequency $\omega_1$ of the incident laser and its direction of propagation $\hat{\mathbf{k}}_1$ are controlled. Given these control variables and plasma parameters, the pump laser can be a superposition of the two electromagnetic (\textit{EM}) eigenmodes. The eigenmode $k_1^-$ with longer wavelength is the \textit{R} wave when $\hat{\mathbf{k}}_1\parallel\mathbf{B}_0$, and it smoothly deforms to the extraordinary (\textit{X}) wave when $\hat{\mathbf{k}}_1\perp\mathbf{B}_0$. On the other hand, the eigenmode $k_1^+$ with shorter wavelength smoothly deforms from the \textit{L} wave to the ordinary (\textit{O}) wave when $\theta_1$, the angle between $\hat{\mathbf{k}}_1$ and $\mathbf{B}_0$, increases from $0^\circ$ to $90^\circ$. Suppose the experiment setup selects one of the eigenmodes, then the wave vector $\mathbf{k}_1$ and polarization $\mathbf{e}_1$ of are fixed. We can then observe frequency $\omega_2$ of the scattered laser using a spectrometer, and select its polarization $\mathbf{e}_2$ using a filter, with both instruments aligned along the $\hat{\mathbf{k}_2}$ direction. If $\omega_3=\omega_1-\omega_2$ and $\mathbf{k}_3=\mathbf{k}_1-\mathbf{k}_2$ correspond to an eigenmode of the plasma, then resonant three-wave scattering can happen, and the spectrometer will display a peak centered at $\omega_2$, whose height is related to the normalized growth rate. 

\begin{figure}[t]
	\includegraphics[angle=0,width=0.5\textwidth]{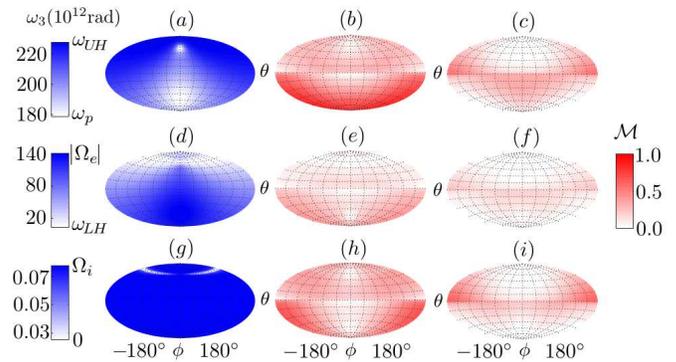}
	\caption{Coherent scattering of an incident laser in the $k_1^+$ eigenmode, propagating in a magnetized cold hydrogen plasma ($\theta_1=30^\circ,\phi_1=0^\circ$). For scattering off the \textit{u}-branch waves, the frequency down shift (a) is between $\omega_p$ and $\omega_{UH}$. The normalized growth rate $\mathcal{M}^+$ of the $k_2^+$ eigenmode (b) is suppressed in polarization-forbidden regions near the equatorial plane ($\theta_2\approx90^\circ$), while the normalized growth rate $\mathcal{M}^-$ of the $k_2^-$ eigenmode (c) is polarization-forbidden in forward ($\theta_2=\theta_1, \phi_2=0^\circ$) and backward ($\theta_2=180^\circ-\theta_1, \phi_2=\pm180^\circ$) directions. When scattered off the \textit{l}-branch waves, the frequency down shift (d) is between $\omega_{LH}$ and $|\Omega_e|$. In addition to polarization forbidden regions, the growth rate $\mathcal{M}^+$ (e) and $\mathcal{M}^-$ (f) are suppressed in interference-forbidden regions where electron and ion scattering cancel (near $\theta_2\approx\theta_1$), as well as in energy-forbidden regions where $\omega_3\approx|\Omega_e|$. Finally, the incident laser can scatter off the \textit{b}-branch waves. The frequency down shift (g) is between zero and $\Omega_i$, and the growth rate $\mathcal{M}^+$ (h) and $\mathcal{M}^-$ (i) are suppressed in polarization forbidden regions, as well as in energy-forbidden regions.}
	\label{fig:ThreeWave}
\end{figure}

As an example, we evaluate the normalized growth rate of a $1.06 \mu$m Nd:glass laser in a magnetized hydrogen plasma, when the incident laser propagates at polar angle $\theta_1=30^\circ$ in the $k_1^+$ eigenmode (Fig.~\ref{fig:ThreeWave}). We take the density of the fully ionized plasma to be $n_0=10^{19} \text{cm}^{-3}$, which is typical for gas jet plasmas. In addition, we take the magnetic field $B_0=8.12$ MG, achievable using existing technologies. In such a plasma, the laser frequency $\omega_1\approx10\omega_p$ and $|\Omega_e|\approx0.8\omega_p$, so the magnetic field plays an important role in coherent Stokes scattering. 
In this two-species cold plasma, three branches of magnetized waves exist, and each branch results in a different angular dependence of the normalized growth rate. 
First, for scattering off the upper (\textit{u}) branch, the frequency down shift (Fig.~\ref{fig:ThreeWave}a) is between $\omega_p$ and $\omega_{UH}$, the upper-hybrid frequency. For the $k_2^+$ eignemode (Fig.~\ref{fig:ThreeWave}b), backscattering is favored while scattering near directions perpendicular to $\hat{\mathbf{k}}_1$, where $\mathbf{e}_1^\dagger\mathbf{e}_2\approx0$, is forbidden. On the contrary, the polarization of $k_2^-$ scattering (Fig.~\ref{fig:ThreeWave}c) is such that forward and backward scattering are forbidden, while perpendicular scattering is allowed. 
Second, when scattered from the lower (\textit{l}) branch, the frequency down shift (Fig.~\ref{fig:ThreeWave}d) is between $|\Omega_e|$ and $\omega_{LH}$, the lower-hybrid frequency. In addition to polarization-forbidden regions near the perpendicular plane, the $k_2^+$ scattering (Fig.~\ref{fig:ThreeWave}e) also encounters special angles where electron and ion scattering exact cancel. Similarly, the $k_2^-$ scattering (Fig.~\ref{fig:ThreeWave}f) have polarization-forbidden regions, as well as regions where scattering from the two species destructively interfere.
Finally, for scattering off the bottom (\textit{b}) branch, the frequency down shift (Fig.~\ref{fig:ThreeWave}g) is between zero and $\Omega_i$. Both $k_2^+$ scattering (Fig.~\ref{fig:ThreeWave}h) and $k_2^-$ scattering (Fig.~\ref{fig:ThreeWave}i) encounter energy forbidden regions near $\theta_2\approx\theta_1$, where plasma waves are energetically too expensive to excite. 
Away from these polarization, interference, and energy forbidden regions where $u_i\gg\Theta$, coherent Stokes scattering from magnetized plasma waves have growth rates comparable to Raman scattering.

\section{\label{sec:amplification}Laser pulse Compression}
While coherent scattering may be an unwanted effect in laser implosion experiments, it can, on the other hand, be utilized to amplify laser pulses beyond what is achievable using other techniques. The current state-of-the-art technique is Chirped Pulse Amplification \cite{Maine88}  (CPA), which can produce intense optical pulses with unfocused intensity on the order of $10^{14} \text{W/cm}^2$ until the damaging threshold of solid gratings is reached \cite{Stuart95,Canova07}. Although laser intensity may be further increased by focusing, the CPA technique is not applicable to shorter wavelength pulses, such as excimer UV lasers \cite{Obenschain96} and free-electron x-ray lasers \cite{Emma10,Ihikawa12}. Since these high-intensity short-wavelength pulses are demanded in many applications, including inertial confinement fusion \cite{Glenzer11,Craxton15} and single molecule imaging \cite{Neutze00,Chapman11}, techniques that can amplify and shorten these pulses are necessary. A promising technique is plasma-based laser pulse compression, using which the pulse intensity can be substantially increased. This technique contemplates using unmagnetized plasma as the gain medium, which supports the Langmuir mode \cite{Malkin99} and the Brillouin mode \cite{Weber13} as mediating waves. However, by magnetizing the plasma medium, not only can we further increase pulse intensity, but we can also extend pulse compression to the soft x-ray regime \cite{Shi17laser}, which was not accessible using previous methods. In this section, we use upper-hybrid wave mediation as an example to demonstrate benefits of applying an external magnetic field in laser pulse compression.

\subsection{Mediation by Upper-Hybrid Wave}
In magnetized plasmas, one of the many waves that can be utilized to mediate laser pulse compression is the upper-hybrid (\textit{UH}) wave. The \textit{UH} wave is the magnetized version of the Langmuir wave, and it is the asymptote of one branch of the electron Bernstein waves in the low temperature limit. The frequency of the \textit{UH} wave is approximately $\omega_{UH}\simeq\sqrt{\omega_p^2+\Omega_e^2}$. Therefore, the external magnetic field partially replaces the role of plasma density in the three-wave resonance condition. In other words, by applying a magnetic field transverse to the direction of laser propagation, the plasma density required to match the resonance condition can be reduced. 

The reduction of requisite plasma density has immediate engineering benefits. First, challenging technology for producing high density plasmas can now be substituted by available technologies for generating strong magnetic fields. 
The plasma density required to compress $1 \mu$m pulses using unmagnetized plasmas is $\sim10^{19} \text{cm}^{-3}$, which is already at the verge of the feasibility of gas jet plasmas. To compress shorter wavelength lasers using unmagnetized plasmas, denser plasma targets, such as foams and aerosol jets \cite{Hay13}, remain to be developed. Allowing the requirement for dense plasmas to be replaced by magnetic fields thus relaxes the engineering challenges. Second, uniformity of the plasma target becomes more maneuverable when magnetic fields supply the resonance frequency. While it is difficult to control the internal plasma density, it is much easier to adjust the external magnetic field to maintain the three-wave resonant condition. Using magnetized plasmas as the gain media thus introduces an extra control variable, using which pulse compression can be tuned.

When the \textit{UH} wave mediates resonant energy transfer between a given pump laser and a given seed pulse, the lower plasma density results in a slower linear growth rate \cite{Shi17laser} $\gamma_0=\sqrt{\omega_3\omega_1}|a_1|/2\gamma_3$, 
where $\gamma_3=\omega_3/\omega_p>1$ is the magnetization factor, defined in Sec.~\ref{sec:scatter}. Other than a smaller growth rate, laser pulse compression mediated by the \textit{UH} wave is similar to Raman compression \cite{Malkin99}. After the linear stage of the amplification, the pump amplitude $a_1$ starts to deplete. At this pump depletion stage, the steep front of the seed pulse keeps on growing, 
whereas the tail of the pulse starts to decay. 
This asymmetric growth of the seed pulse results in an effective compression of the pulse duration. After the seed pulse transit the entire length of the pump laser, it emerges as an amplified pulse with a shortened duration. Since \textit{UH} wave mediation has smaller growth rate, it takes longer time, and equivalently, longer pump laser and plasma length, to achieve the same compression of the seed pulse. 

\subsection{Limiting Effects}
Although the amplification rate is reduced for \textit{UH} mediation, the growth rates of competing instabilities are reduced more. Therefore, one can use longer pumps than allowed in unmagnetized plasmas to amplify seed pulses when the media become magnetized. For example, one of the most competitive instabilities is the modulational instability of the seed pulse, whose growth rate \cite{Shi17laser} $\gamma_M=\omega_3^2|a_2|^2/8\omega_1\gamma_3^{2}$
is reduced by an additional factor of $\gamma_3>1$. After a few exponentiations, the modulational instability causes the leading spike of the pulse to break up, and thereof limits the allowable pulse amplification time $t_M$. Allowing for the possibility that subdominant spikes in the pulse train may further grow \cite{Malkin14,Barth16}, we may estimate a lower bound \cite{Shi17laser} $t_M\propto\omega_3^{-1}\gamma_3^{4/3}$. Notice that the exponent of $\gamma_3$ is larger than one. In fact, this exponent is larger than $3/2$ in particle-in-cell simulations \cite{Jia17}. The net consequence of a smaller amplification rate but a longer allowable amplification time is thus a higher achievable pulse intensity when we magnetize the plasma medium.

In addition to relatively suppressing competing instabilities, replacing plasma density with magnetic fields also reduces wave damping, a sink of wave energy that could otherwise be used to amplify the laser pulse. Since collisional damping rates of \textit{EM} waves are $\nu_{1,2}\simeq\nu_{ei}\omega_p^2/2\omega_{1,2}^2$, where $\nu_{ei}=n_eZ^2e^4\Lambda/(4\pi\epsilon_0)^2m_e^2v^3$ is the electron-ion collision frequency, the damping rates of \textit{EM} waves scale with plasma density as $\nu_{1,2}\propto n_e^2$, as expected of two-body collisions. Therefore, when plasma density is replaced by magnetic fields, collisional damping of \textit{EM} waves can be substantially reduced. In addition to collisional damping, the plasma wave also suffers from collisionless damping. Although linear collisionless damping exactly vanishes when the plasma wave propagates perpendicular to the magnetic field \cite{Stix92}, nonlinear mechanisms, such as stochastic heating \cite{Karney78,Karney79} and surfatron acceleration \cite{Sagdeev73,Dawson83}, can still damp the plasma wave. Since collisionless damping is due to phase mixing, its rate scales as $\nu_3\propto n_e$. Therefore, collisionless damping is also reduced when we magnetized the plasma medium.

While moderate magnetic fields improve performance of laser pulse compression, it is not favorable to impose a magnetic field that is too strong due to wakefield generation. Laser wakefields are generated when the ponderomotive force of the laser pulse expels electrons to form plasma bubbles. When plasma density is reduced, wakefields can thus be excited more easily. Moreover, as the magnetization factor increases, the spectra of wakefields broaden \cite{Holkundkar11} and the electromagnetic components of wakefields enlarge \cite{Jia17}. Consequently, magnetized wakefields contain larger degrees of freedom, which allow them to partition a larger fraction of the total energy during nonlinear interactions. Energy in the wakefields can thereafter be transferred irreversibly to energize electrons. In addition to wakefield acceleration in the direction of laser propagation, electrons are also accelerated in the perpendicular direction due to the magnetic field, which allows them to enter and leave plasma bubbles in the transverse direction. Therefore, when magnetic fields increase beyond the optimal value, magnetized wakefields inhibit further growth of the laser pulse.

Fortunately, the vulnerability due to electromagnetic wakefield generation may be compensated by the resilience of magnetized plasmas to wavebreaking. Wavebreaking of the plasma wave is what limits the viable pump laser intensity in unmagnetized plasmas. When the pump intensity exceeds the wavebreaking threshold, the excited plasma wave becomes so strong that the quivering electrons outrun the phase velocity of the wave, leading to collapse of the plasma wave envelope. However, when the plasma density is replaced by a magnetic field, not only is the plasma wave electric field reduced, but the magnetic field also provides an additional restoring force. Therefore, the plasma wave remains coherent even when the pump laser exceeds the wavebreaking threshold \cite{Jia17}. Until a larger phase mixing threshold is reached, we can use more intense pump lasers than allowed in unmagnetized plasmas to amplify the seed pulse to higher intensity.

\subsection{Validation Using PIC simulations}
\begin{figure}[t]
	\includegraphics[angle=0,width=0.44\textwidth]{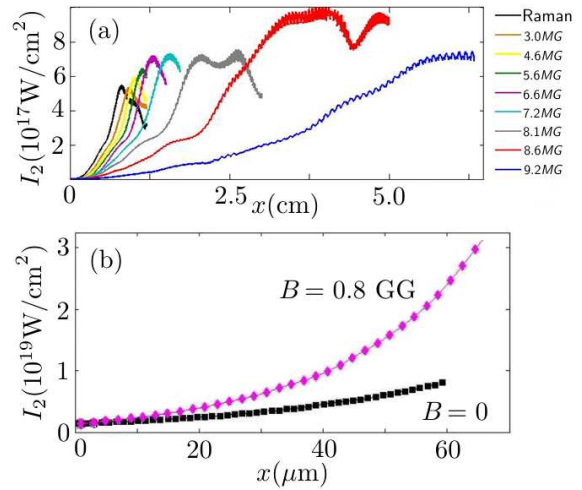}
	\caption{Applying a transverse magnetic field improves the performance of plasma-based laser pulse compression, as shown here using PIC simulations. For a $1\mu$m optical pulse (a), using a longer plasma and an optimal magnetic field (red line), we can double the pulse intensity achievable with unmagnetized Raman (black line). For a shorter-wavelength $10$ nm x-ray pulse (b), replacing plasma density with a transverse magnetic field on giga-Gauss scale alleviates strong damping. Consequently, magnetized pulse compression becomes possible (purple), while unmagnetized amplification can barely work (black).}
	\label{fig:PIC}
\end{figure}

The prediction that applying a moderate magnetic field improves the performance of laser pulse compression has been verified using particle-in-cell (PIC) simulations \cite{Jia17}. In a set of one-dimensional PIC simulations, we use a $1.0 \mu$m pump laser, with initial intensity $I_{10}=3.5\times10^{14} \text{W/cm}^2$, to compress a counter-propagating $1.1 \mu$m seed pulse, with initial intensity $I_{20}=1.8\times10^{13} \text{W/cm}^2$ and initial duration $\Delta t_{20}=33$ fs. Given the pump and the seed lasers, we apply a magnetic field transverse to the direction of laser propagation, and reduce the plasma density accordingly to maintain the resonance condition (Fig.~\ref{fig:PIC}a). When there is no magnetic field (black line), pulse compression is mediated by Raman backscattering. After the initial exponential growth, the seed pulse enters the nonlinear compression stage, until its intensity is saturated at $I_{2}\approx5.5\times10^{17} \text{W/cm}^2$ due to the modulational instability. As we increase the magnetic field (color lines), the growth becomes slower, but the saturation is delayed. The net consequence is that the attainable final pulse intensity increases with the magnetic field, until an optimal field $B\approx 8.6$ MG is reached (red line), where the pulse intensity is about twice of what is achievable using Raman compression. When a stronger magnetic field is applied (blue line), the seed pulse loses a substantial amount of energy to the wakefield, which inhibits further improvements of the pulse intensity.

In addition to improving the performance in the optical regime, applying a magnetic field enables compression of short-wavelength pulses that cannot be compressed using unmagnetized plasmas. For example, a $10$ nm soft x-ray laser is at the verge of what can be compressed using Raman compression \cite{Malkin07}. At even shorter wavelength, collisional damping becomes too strong. The total damping could have been alleviated by increasing the plasma temperature, if it were not due to collisionless damping, which increases with the plasma temperature. Therefore, the operation window in the plasma parameter space is almost closed \cite{Shi17laser}. In one-dimensional PIC simulations \cite{Jia}, the $11$ nm seed pulse, whose initial intensity $I_{20}=1.4\times10^{18} \text{W/cm}^2$ and initial duration $\Delta t_{20}=1.5$ fs, barely grows (Fig.~\ref{fig:PIC}b, black), when the seed pulse transits a pump laser with $I_{10}=1.4\times10^{18} \text{W/cm}^2$. However, by replacing plasma density with a $0.8$ GG magnetic field, the effective growth rate becomes much larger (purple). This is because although the undamped growth rate $\gamma_0\propto n_e^{1/2}$ is reduced in lower density plasmas, the collisionless damping $\nu_{3}\propto n_e$ and the collisional damping $\nu_{1,2}\propto n_e^2$ are reduced more substantially. Therefore, faster effective growth is possible when we magnetized the plasma medium, using which compression of soft x-ray pulses beyond the reach of previous methods becomes possible.

\section{\label{sec:propagation}Light propagation in QED regime}
While mega-Gauss magnetic fields introduce new phenomena and applications in the classical regime, an even stronger giga-Gauss magnetic field may already enable us to probe relativistic quantum physics in the QED regime \cite{Shi16QED}. To see when relativistic quantum effects are important, we can compare energy scales in the system. The typical energy scales of plasmas are thermal energy $k_BT$, Fermi energy $\epsilon_F$, plasma energy $\epsilon_p=\omega_p\hbar$ and gyro energy $\epsilon_g=\Omega_e\hbar$. The energy scales of the fields are electric energy $\epsilon_E=\sqrt{eEc\hbar}$, magnetic energy $\epsilon_B=\sqrt{eBc^2\hbar}$, photon energy $\epsilon_\gamma=\omega_\gamma\hbar$ and ponderomotive energy $U_p$. Relativistic effects are important when any of these energy scales becomes comparable to the electron rest energy $m_ec^2$, and quantum effects are important whenever non-thermal energy $\epsilon_*$ dominates the thermal energy. An example where both relativistic and quantum effects are important is magnetospheres of x-ray pulsars. The typical magnetic field $B\sim10^{12}$ G corresponds to $\epsilon_B\sim100$ KeV. Since $\epsilon_B$ is comparable to $m_ec^2\approx511$ KeV, relativistic effects are important. At the same time, $\epsilon_B$ is much higher than the thermal energy $k_BT\sim10$ KeV, which makes quantum effects important. Although magnetic fields of such strength are not yet available in laboratory, it turns out that small relativistic quantum corrections may already be observable in giga-Gauss magnetic field through Faraday rotation.

\subsection{Dispersion Relation in QED plasmas}
To interpret signals from strongly magnetized plasmas, it is imperative that we understand how light propagates in the relativistic-quantum regime. Propagation of small amplitude waves, such as photons, is governed by the linear dispersion relation. The dispersion relation can be derived from the linearized momentum space wave equation, which can be written in a Lorentz-invariant and gauge-invariant form $(k^{\mu}k^{\nu}-k^2g^{\mu\nu}+\hat{\Sigma}^{\mu\nu})A_{\nu}=0$, where $k^{\mu}$ is the wave 4-momentum and $g^{\mu\nu}$ is the Minkowski metric. The response tensor $\hat{\Sigma}^{\mu\nu}$, which satisfies the Ward-Takahashi identity $\hat{\Sigma}^{\mu\nu}k_{\nu}=k_{\mu}\hat{\Sigma}^{\mu\nu}=0$, describes reactions felt by photons, as they polarize both the plasma medium and the vacuum when they propagate. Since the dispersion relation is gauge invariant, we can, for example, choose the temporal gauge $A_0=0$. In this gauge, the linear dispersion relation becomes
\begin{equation}\label{eq:MagDisp}
\det\!\left(\!\begin{array}{ccc}
\omega^2\!-\!k_{\parallel}^2\!+\!\hat{\Sigma}^{11}\!&\!\hat{\Sigma}^{12}\!&\!k_\perp k_\parallel\!+\!\hat{\Sigma}^{13}\!\\
\hat{\Sigma}^{21}\!&\!\omega^2\!-\!\bm{k}^2\!+\!\hat{\Sigma}^{22}\!&\!\hat{\Sigma}^{23}\!\\
k_\perp k_\parallel\!+\!\hat{\Sigma}^{31}\!&\!\hat{\Sigma}^{32}\!&\!\omega^2\!-\!k_\perp^2\!+\!\hat{\Sigma}^{33}\!
\end{array}\!\right)\!=\!0.
\end{equation}
Here, we have chosen a coordinate system in which the wave 4-momentum $k^{\mu}=(\omega,k_\perp,0,k_\parallel)$, where we have used the natural unit $\hbar=c=\epsilon_0=1$. In this form, it is easy to see that the spatial components of the response tensor $\hat{\Sigma}^{ij}=\omega^2\chi^{ij}$ is related to the linear susceptibility.  

While the dispersion relation is formally identical to that in the classical regime, relativistic-quantum physics are encoded in the response tensor. As an example, let us consider the response tensor in a magnetized scalar-QED plasma in its ground state, when charged particle responses dominate the vaccum response. In the coordinate system where $\mathbf{B}_0=(0,0,B_0)$ and $\mathbf{k}=(k_\perp,0,k_\parallel)$, the diagonal components of the response tensor are \cite{Shi16QED}
\begin{eqnarray}
\nonumber
\hat{\Sigma}^{11}&=&-\frac{m\omega_p^2}{2m_0}\sum_{\varsigma=\pm 1}\kappa_{\varsigma}^2\mathbb{K}_\varsigma^{(1)},\\
\nonumber
\hat{\Sigma}^{22}&=&\hat{\Sigma}^{11}-\frac{m\omega_p^2}{2m_0}\sum_{\varsigma=\pm 1}\kappa_{\perp}^2\Big(\mathbb{K}_\varsigma^{(0)}-2\mathbb{K}_\varsigma^{(1)}\Big),\\
\hat{\Sigma}^{33}&=&-\frac{m\omega_p^2}{m_0}\Big(1+\frac{1}{2}\sum_{\varsigma=\pm 1}\kappa_{\parallel}^2\mathbb{K}_\varsigma^{(0)}\Big),
\end{eqnarray}
where summation over charged species is implied. In the above formulas, $\omega_p^2=e^2n_0/m$ is the plasma frequency, and $m_0=\sqrt{m^2+eB_0}$ is the shifted ground state mass. The summation over $\varsigma=\pm 1$ corresponds to the summation of the $s$-channel and the $t$-channel Feynman diagrams. Inside the summations, $\kappa^{\mu}=l_Bk^{\mu}/2$ is the wave 4-momentum normalized by the magnetic length $l_B=\sqrt{2/eB_0}$. The kernel of the propagator $\kappa_\varsigma^2:=\kappa_0^2-\kappa_\parallel^2+\varsigma\varrho_0\kappa_0$, where $\varrho^{\mu}=l_B(m_0,0,0,0)$ is the normalized 4-momentum of particles in the ground state. For conciseness, we denote $\mathbb{K}_\varsigma^{(n)}:= \mathbb{K}(\kappa_\varsigma^2-n,\kappa_\perp^2)$, where the $\mathbb{K}$-function is related to the confluent hypergeometric function ${}_1F_1(a;b;z)$ by $\mathbb{K}(x,z):={}_1F_1(1;1-x;-z)/x$. Similarly, the off-diagonal components of the response tensor can be expressed in terms of the $\mathbb{K}$-function as \cite{Shi16QED}
\begin{eqnarray}\label{MagPol}
\nonumber
\hat{\Sigma}^{12}&=&-\hat{\Sigma}^{21}=-i\frac{m\omega_p^2}{2m_0}\sum_{\varsigma=\pm 1} \varsigma\kappa_{\varsigma}^2\Big(\mathbb{K}_{\varsigma}^{(1)}-\mathbb{K}_\varsigma^{(0)}\Big),\\
\nonumber
\hat{\Sigma}^{23}&=&-\hat{\Sigma}^{32}=+i\frac{m\omega_p^2}{2m_0}\sum_{\varsigma=\pm 1} \varsigma\kappa_{\perp}\kappa_{\parallel}\Big(\mathbb{K}_{\varsigma}^{(1)}-\mathbb{K}_\varsigma^{(0)}\Big),\\
\nonumber
\hat{\Sigma}^{31}&=&+\hat{\Sigma}^{13}=-\frac{m\omega_p^2}{2m_0}\sum_{\varsigma=\pm 1}\kappa_{\perp}\kappa_{\parallel}\mathbb{K}_{\varsigma}^{(1)} .
\end{eqnarray}
The confluent hypergeometric function arises when we compute the response tensor by summing over all transitions between relativistic Landau levels. 

\subsection{Modifications to Faraday Rotation}
As a special case, consider photon propagation parallel to the magnetic field, in which case relativistic-quantum effects modify the Faraday rotation. For exact parallel propagation, $k_\perp=0$, and the $\mathbb{K}$-functions take special values $\mathbb{K}_{\varsigma}^{(n)}=1/(\kappa_{\varsigma}^2-n)$. Substituting these special values into the dispersion relation, it is straightforwards to show that the two electromagnetic eigenmodes are the \textit{R} wave, which satisfies $n^2=R$, and the \textit{L} wave, which satisfies $n^2=L$. Here, $n^2=c^2k_\parallel^2/\omega^2$ is the refractive index, and the permittivities \cite{Shi16QED}
\begin{eqnarray}
R&=&1-\sum_{s}\frac{m_s\omega_{ps}^2}{m_{s0}\omega^2}\frac{\omega^2-k_\parallel^2-2m_{s0}\omega}{\omega^2-k_\parallel^2- 2(m_{s0}\omega+m_s\Omega_s)},\\
L&=&1-\sum_{s}\frac{m_s\omega_{ps}^2}{m_{s0}\omega^2}\frac{\omega^2-k_\parallel^2+2m_{s0}\omega}{\omega^2-k_\parallel^2+ 2(m_{s0}\omega-m_s\Omega_s)}.
\end{eqnarray} 
In the classical limit $\omega,k_\parallel,\Omega\ll m$, the above formulas recover the classical results. In the opposite limit, relativistic-quantum effects may substantially modify the dispersion relation.

One way to observe relativistic-quantum modifications is to measure Faraday rotations using a set of linearly polarized lasers, propagating together along the magnetic field in the $z$-direction. Due to the difference in phase velocities of the \textit{R} wave and the \textit{L} wave, the polarization vector of each linearly polarized laser rotates at a rate
\begin{equation}
\frac{d\psi}{d\zeta}=\pi\Delta n,
\end{equation}
where $\psi$ is the polarization angle, $\zeta=z\omega/2\pi c$ is the distance of propagation normalized by the vacuum wavelength $\lambda_0$, and $\Delta n=n_L-n_R$ is the difference in refractive indexes between the \textit{L} wave and the \textit{R} wave of the same frequency. In an electron-ion plasma, since $m_i\gg m_e$, the dominant contribution comes from electrons. Keeping only electron terms in the dispersion relations, the refractive index
\begin{eqnarray}
\label{eq:nRL}
\nonumber
n_{R/L}^2&=&1-\frac{m\Omega}{\omega^2}-\frac{m\omega_p^2}{2m_0\omega^2}\mp\frac{m_0}{\omega} \\
&\pm&\sqrt{\Big(\frac{m\Omega}{\omega^2}+\frac{m\omega_p^2}{2m_0\omega^2}\pm\frac{m_0}{\omega}\Big)^2\mp\frac{2m\omega_p^2}{\omega^3}},
\end{eqnarray}
where the upper signs correspond to the \textit{R} wave and the lower signs correspond to the \textit{L} wave. It is straightforward to check that in the classical limit $\omega,\omega_p,\Omega\ll m$, the above formulas recover the classical results. As a side remark, notice that in electron-positron plasmas with charge conjugation symmetry, the Faraday rotation remains identically zero as in the classical case.

Although relativistic-quantum modifications are still small in giga-Gauss magnetic fields, they are boosted near the cutoff frequency of the \textit{R} wave, where Faraday rotation is maximized. Suppose we measure Faraday rotations by passing multiple lasers of slightly different frequencies through the same plasma, then the relativistic-quantum formula predicts a different frequency dependence than given by the classical formula. To see the difference, one can subtract measured data from classical predictions, and plot $\Delta\psi$ as function of laser frequencies (Fig.~\ref{fig:Faraday}). 
For example, in a gas jet plasma with density $n_e=10^{19} \text{cm}^{-3}$, a magnetic field $B_0=10^{8}$ G results in a difference of $\sim 1^\circ/\lambda_0$ when the laser frequency approaches the \textit{R}-wave cutoff  $\sim 1.16$ eV (red line). This discrepancy can be resolved if the measurement uncertainty is $\lesssim1.5\%$ at the classical cutoff, and $\lesssim 15$ ppm at $\sim0.1$ eV above the cutoff. 
In a stronger magnetic field $B_0=10^{9}$ G, the difference is as large as $\sim 10^\circ/\lambda_0$ near the cutoff $\sim 11.5$ eV (blue line). This discrepancy can be resolved if measurement uncertainty is $\lesssim67\%$ at the classical cutoff, and $\lesssim0.13\%$ at $\sim0.1$ eV above the cutoff.
While corrections introduced by a 0.1 GG magnetic field is unlikely to be measurable, much larger corrections introduced by giga-Gauss magnetic fields might be discernible from noise and inhomogeneities, after accumulating the difference by a few vacuum wavelengths.

%

\section{\label{sec:conclusion}Summary and Discussion}
In this paper, we review three research directions, addressing challenges and opportunities when strong magnetic fields become available. First, we provide a convenient formula for resonant three-wave coupling [Eq.~(\ref{eq:coupling})] in magnetized plasmas. Using hydrogen plasma as an example, we identified special angles where the scattering is polarization, interference, and energy forbidden (Fig.~\ref{fig:ThreeWave}). Away from these forbidden angles, coherent laser scattering in a magnetized plasma has growth rate comparable to Raman scattering. Consequently, magnetic fields may be applied to either suppress or enhance laser scattering at selected angles. Further analysis of thermal effects and wave damping in the presence of multiple lasers may enable optimization of laser-plasma coupling in magnetized laser implosion experiments.

Second, we analyze benefits of applying magnetic fields in laser pulse compression. In addition to relaxing engineering constraints, substituting plasma density with a moderate magnetic field suppresses competing instabilities and reduces wave damping. These improvements enable us to use the magnetic field as an extra control variable to optimize the performance of $1 \mu$m lasers pulse compression (Fig.~\ref{fig:PIC}a). Moreover, using upper-hybrid mediation, compression of soft x-ray pulses beyond the reach of unmagnetized schemes now becomes possible (Fig.~\ref{fig:PIC}b). These results, obtained from simple analytical estimations and 1D PIC simulations, remain to be verified by more comprehensive simulations and ultimately by experiments. In addition to upper-hybrid mediation, the possibilities of using other magnetized plasma waves, such as Alfv\'en waves, hybrid waves, and Bernstein waves, to mediate pulse compression remain to be analyzed.  

\begin{figure}[t]
	\includegraphics[angle=0,width=0.35\textwidth]{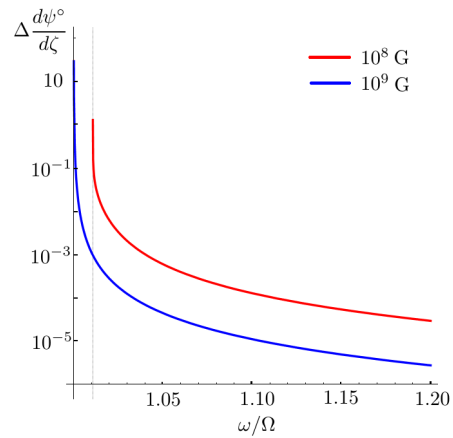}
	\caption{Deviations of Faraday rotation from classical predictions can be used to measure relativistic-quantum corrections. In a gas jet plasma with density $n_e=10^{19} \text{cm}^{-3}$, a 0.1 GG magnetic field (red) leads to a deviation $\Delta\psi$ of $\sim 1^\circ$ after the laser, whose frequency is near the \textit{R}-wave cutoff, propagates by a vacuum wavelength $\lambda_0$. In a stronger 1 GG magnetic field (blue), a deviation as large as $\sim 10^\circ/\lambda_0$ may be observed using a laser whose frequency is slightly above the cutoff. Notice that the deviations fall precipitously when laser frequencies are above the classical cutoff. Therefore, for relativistic-quantum effects to be measurable, the laser frequency must be sufficiently close to the cutoff.}
	\label{fig:Faraday}
\end{figure}

Finally, we speculate experimental possibilities of entering relativistic-quantum regime using giga-Gauss magnetic fields, which are at the cusp of current feasibility. Although such magnetic fields are still much smaller than the Schwinger field, relativistic-quantum modifications may already be observable, albeit very challenging, using Faraday rotation (Fig.~\ref{fig:Faraday}). While results in cold scalar-QED plasmas may be instructive, laboratory plasmas are typically made of spinor particles at finite temperature. Therefore, spin and thermal effects on experimental observables remain to be studied. Beyond the perturbative regime, it might be possible to use a combination of a strong magnetic field and an intense laser to probe relativistic-quantum physics that could be probed by neither the magnetic field nor the laser alone. For this purpose, developing numerical schemes that can simulate relativistic-quantum plasmas will be indispensable.

\begin{acknowledgments} 
Special thanks to Dr.~Qing Jia for carrying out PIC simulations on PPPL's Research Computing Center using the EPOCH code, and providing us with Fig.~\ref{fig:PIC} in this paper. This research is supported by NNSA Grant No. DE-NA0002948, AFOSR Grant No. FA9550-15-1-0391, and DOE Grant No. DEAC02-09CH11466.
\end{acknowledgments}

%
%



\begin{thebibliography}{61}%
	\makeatletter
	\providecommand \@ifxundefined [1]{%
		\@ifx{#1\undefined}
	}%
	\providecommand \@ifnum [1]{%
		\ifnum #1\expandafter \@firstoftwo
		\else \expandafter \@secondoftwo
		\fi
	}%
	\providecommand \@ifx [1]{%
		\ifx #1\expandafter \@firstoftwo
		\else \expandafter \@secondoftwo
		\fi
	}%
	\providecommand \natexlab [1]{#1}%
	\providecommand \enquote  [1]{``#1''}%
	\providecommand \bibnamefont  [1]{#1}%
	\providecommand \bibfnamefont [1]{#1}%
	\providecommand \citenamefont [1]{#1}%
	\providecommand \href@noop [0]{\@secondoftwo}%
	\providecommand \href [0]{\begingroup \@sanitize@url \@href}%
	\providecommand \@href[1]{\@@startlink{#1}\@@href}%
	\providecommand \@@href[1]{\endgroup#1\@@endlink}%
	\providecommand \@sanitize@url [0]{\catcode `\\12\catcode `\$12\catcode
		`\&12\catcode `\#12\catcode `\^12\catcode `\_12\catcode `\%12\relax}%
	\providecommand \@@startlink[1]{}%
	\providecommand \@@endlink[0]{}%
	\providecommand \url  [0]{\begingroup\@sanitize@url \@url }%
	\providecommand \@url [1]{\endgroup\@href {#1}{\urlprefix }}%
	\providecommand \urlprefix  [0]{URL }%
	\providecommand \Eprint [0]{\href }%
	\providecommand \doibase [0]{http://dx.doi.org/}%
	\providecommand \selectlanguage [0]{\@gobble}%
	\providecommand \bibinfo  [0]{\@secondoftwo}%
	\providecommand \bibfield  [0]{\@secondoftwo}%
	\providecommand \translation [1]{[#1]}%
	\providecommand \BibitemOpen [0]{}%
	\providecommand \bibitemStop [0]{}%
	\providecommand \bibitemNoStop [0]{.\EOS\space}%
	\providecommand \EOS [0]{\spacefactor3000\relax}%
	\providecommand \BibitemShut  [1]{\csname bibitem#1\endcsname}%
	\let\auto@bib@innerbib\@empty
	\bibitem [{\citenamefont {Igumenshchev}\ \emph {et~al.}(2014)\citenamefont
		{Igumenshchev}, \citenamefont {Zylstra}, \citenamefont {Li}, \citenamefont
		{Nilson}, \citenamefont {Goncharov},\ and\ \citenamefont
		{Petrasso}}]{Igumenshchev14}%
	\BibitemOpen
	\bibfield  {author} {\bibinfo {author} {\bibfnamefont {I.}~\bibnamefont
			{Igumenshchev}}, \bibinfo {author} {\bibfnamefont {A.}~\bibnamefont
			{Zylstra}}, \bibinfo {author} {\bibfnamefont {C.}~\bibnamefont {Li}},
		\bibinfo {author} {\bibfnamefont {P.}~\bibnamefont {Nilson}}, \bibinfo
		{author} {\bibfnamefont {V.}~\bibnamefont {Goncharov}}, \ and\ \bibinfo
		{author} {\bibfnamefont {R.}~\bibnamefont {Petrasso}},\ }\bibfield  {title}
	{\enquote {\bibinfo {title} {Self-generated magnetic fields in direct-drive
				implosion experiments},}\ }\href@noop {} {\bibfield  {journal} {\bibinfo
			{journal} {Phys. Plasmas}\ }\textbf {\bibinfo {volume} {21}},\ \bibinfo
		{pages} {062707} (\bibinfo {year} {2014})}\BibitemShut {NoStop}%
	\bibitem [{\citenamefont {Gotchev}\ \emph {et~al.}(2009)\citenamefont
		{Gotchev}, \citenamefont {Chang}, \citenamefont {Knauer}, \citenamefont
		{Meyerhofer}, \citenamefont {Polomarov}, \citenamefont {Frenje},
		\citenamefont {Li}, \citenamefont {Manuel}, \citenamefont {Petrasso},
		\citenamefont {Rygg} \emph {et~al.}}]{Gotchev09}%
	\BibitemOpen
	\bibfield  {author} {\bibinfo {author} {\bibfnamefont {O.}~\bibnamefont
			{Gotchev}}, \bibinfo {author} {\bibfnamefont {P.}~\bibnamefont {Chang}},
		\bibinfo {author} {\bibfnamefont {J.}~\bibnamefont {Knauer}}, \bibinfo
		{author} {\bibfnamefont {D.}~\bibnamefont {Meyerhofer}}, \bibinfo {author}
		{\bibfnamefont {O.}~\bibnamefont {Polomarov}}, \bibinfo {author}
		{\bibfnamefont {J.}~\bibnamefont {Frenje}}, \bibinfo {author} {\bibfnamefont
			{C.}~\bibnamefont {Li}}, \bibinfo {author} {\bibfnamefont {M.-E.}\
			\bibnamefont {Manuel}}, \bibinfo {author} {\bibfnamefont {R.}~\bibnamefont
			{Petrasso}}, \bibinfo {author} {\bibfnamefont {J.}~\bibnamefont {Rygg}},
		\emph {et~al.},\ }\bibfield  {title} {\enquote {\bibinfo {title}
			{Laser-driven magnetic-flux compression in high-energy-density plasmas},}\
	}\href@noop {} {\bibfield  {journal} {\bibinfo  {journal} {Phys. Rev. Lett.}\
		}\textbf {\bibinfo {volume} {103}},\ \bibinfo {pages} {215004} (\bibinfo
		{year} {2009})}\BibitemShut {NoStop}%
	\bibitem [{\citenamefont {Knauer}\ \emph {et~al.}(2010)\citenamefont {Knauer},
		\citenamefont {Gotchev}, \citenamefont {Chang}, \citenamefont {Meyerhofer},
		\citenamefont {Polomarov}, \citenamefont {Betti}, \citenamefont {Frenje},
		\citenamefont {Li}, \citenamefont {Manuel}, \citenamefont {Petrasso} \emph
		{et~al.}}]{Knauer10}%
	\BibitemOpen
	\bibfield  {author} {\bibinfo {author} {\bibfnamefont {J.}~\bibnamefont
			{Knauer}}, \bibinfo {author} {\bibfnamefont {O.}~\bibnamefont {Gotchev}},
		\bibinfo {author} {\bibfnamefont {P.}~\bibnamefont {Chang}}, \bibinfo
		{author} {\bibfnamefont {D.}~\bibnamefont {Meyerhofer}}, \bibinfo {author}
		{\bibfnamefont {O.}~\bibnamefont {Polomarov}}, \bibinfo {author}
		{\bibfnamefont {R.}~\bibnamefont {Betti}}, \bibinfo {author} {\bibfnamefont
			{J.}~\bibnamefont {Frenje}}, \bibinfo {author} {\bibfnamefont
			{C.}~\bibnamefont {Li}}, \bibinfo {author} {\bibfnamefont {M.-E.}\
			\bibnamefont {Manuel}}, \bibinfo {author} {\bibfnamefont {R.}~\bibnamefont
			{Petrasso}},  \emph {et~al.},\ }\bibfield  {title} {\enquote {\bibinfo
			{title} {Compressing magnetic fields with high-energy lasers},}\ }\href@noop
	{} {\bibfield  {journal} {\bibinfo  {journal} {Phys. Plasmas}\ }\textbf
		{\bibinfo {volume} {17}},\ \bibinfo {pages} {056318} (\bibinfo {year}
		{2010})}\BibitemShut {NoStop}%
	\bibitem [{\citenamefont {Fujioka}\ \emph {et~al.}(2013)\citenamefont
		{Fujioka}, \citenamefont {Zhang}, \citenamefont {Ishihara}, \citenamefont
		{Shigemori}, \citenamefont {Hironaka}, \citenamefont {Johzaki}, \citenamefont
		{Sunahara}, \citenamefont {Yamamoto}, \citenamefont {Nakashima},
		\citenamefont {Watanabe} \emph {et~al.}}]{Fujioka13}%
	\BibitemOpen
	\bibfield  {author} {\bibinfo {author} {\bibfnamefont {S.}~\bibnamefont
			{Fujioka}}, \bibinfo {author} {\bibfnamefont {Z.}~\bibnamefont {Zhang}},
		\bibinfo {author} {\bibfnamefont {K.}~\bibnamefont {Ishihara}}, \bibinfo
		{author} {\bibfnamefont {K.}~\bibnamefont {Shigemori}}, \bibinfo {author}
		{\bibfnamefont {Y.}~\bibnamefont {Hironaka}}, \bibinfo {author}
		{\bibfnamefont {T.}~\bibnamefont {Johzaki}}, \bibinfo {author} {\bibfnamefont
			{A.}~\bibnamefont {Sunahara}}, \bibinfo {author} {\bibfnamefont
			{N.}~\bibnamefont {Yamamoto}}, \bibinfo {author} {\bibfnamefont
			{H.}~\bibnamefont {Nakashima}}, \bibinfo {author} {\bibfnamefont
			{T.}~\bibnamefont {Watanabe}},  \emph {et~al.},\ }\bibfield  {title}
	{\enquote {\bibinfo {title} {Kilotesla magnetic field due to a capacitor-coil
				target driven by high power laser},}\ }\href@noop {} {\bibfield  {journal}
		{\bibinfo  {journal} {Sci. Rep.}\ }\textbf {\bibinfo {volume} {3}},\ \bibinfo
		{pages} {1170} (\bibinfo {year} {2013})}\BibitemShut {NoStop}%
	\bibitem [{\citenamefont {Santos}\ \emph {et~al.}(2015)\citenamefont {Santos},
		\citenamefont {Bailly-Grandvaux}, \citenamefont {Giuffrida}, \citenamefont
		{Forestier-Colleoni}, \citenamefont {Fujioka}, \citenamefont {Zhang},
		\citenamefont {Korneev}, \citenamefont {Bouillaud}, \citenamefont {Dorard},
		\citenamefont {Batani} \emph {et~al.}}]{Santos15}%
	\BibitemOpen
	\bibfield  {author} {\bibinfo {author} {\bibfnamefont {J.}~\bibnamefont
			{Santos}}, \bibinfo {author} {\bibfnamefont {M.}~\bibnamefont
			{Bailly-Grandvaux}}, \bibinfo {author} {\bibfnamefont {L.}~\bibnamefont
			{Giuffrida}}, \bibinfo {author} {\bibfnamefont {P.}~\bibnamefont
			{Forestier-Colleoni}}, \bibinfo {author} {\bibfnamefont {S.}~\bibnamefont
			{Fujioka}}, \bibinfo {author} {\bibfnamefont {Z.}~\bibnamefont {Zhang}},
		\bibinfo {author} {\bibfnamefont {P.}~\bibnamefont {Korneev}}, \bibinfo
		{author} {\bibfnamefont {R.}~\bibnamefont {Bouillaud}}, \bibinfo {author}
		{\bibfnamefont {S.}~\bibnamefont {Dorard}}, \bibinfo {author} {\bibfnamefont
			{D.}~\bibnamefont {Batani}},  \emph {et~al.},\ }\bibfield  {title} {\enquote
		{\bibinfo {title} {Laser-driven platform for generation and characterization
				of strong quasi-static magnetic fields},}\ }\href@noop {} {\bibfield
		{journal} {\bibinfo  {journal} {New J. Phys.}\ }\textbf {\bibinfo {volume}
			{17}},\ \bibinfo {pages} {083051} (\bibinfo {year} {2015})}\BibitemShut
	{NoStop}%
	\bibitem [{\citenamefont {Goyon}\ \emph {et~al.}(2017)\citenamefont {Goyon},
		\citenamefont {Pollock}, \citenamefont {Turnbull}, \citenamefont {Hazi},
		\citenamefont {Divol}, \citenamefont {Farmer}, \citenamefont {Haberberger},
		\citenamefont {Javedani}, \citenamefont {Johnson}, \citenamefont {Kemp} \emph
		{et~al.}}]{Goyon17}%
	\BibitemOpen
	\bibfield  {author} {\bibinfo {author} {\bibfnamefont {C.}~\bibnamefont
			{Goyon}}, \bibinfo {author} {\bibfnamefont {B.}~\bibnamefont {Pollock}},
		\bibinfo {author} {\bibfnamefont {D.}~\bibnamefont {Turnbull}}, \bibinfo
		{author} {\bibfnamefont {A.}~\bibnamefont {Hazi}}, \bibinfo {author}
		{\bibfnamefont {L.}~\bibnamefont {Divol}}, \bibinfo {author} {\bibfnamefont
			{W.}~\bibnamefont {Farmer}}, \bibinfo {author} {\bibfnamefont
			{D.}~\bibnamefont {Haberberger}}, \bibinfo {author} {\bibfnamefont
			{J.}~\bibnamefont {Javedani}}, \bibinfo {author} {\bibfnamefont
			{A.}~\bibnamefont {Johnson}}, \bibinfo {author} {\bibfnamefont
			{A.}~\bibnamefont {Kemp}},  \emph {et~al.},\ }\bibfield  {title} {\enquote
		{\bibinfo {title} {Ultrafast probing of magnetic field growth inside a
				laser-driven solenoid},}\ }\href@noop {} {\bibfield  {journal} {\bibinfo
			{journal} {Phys. Rev. E}\ }\textbf {\bibinfo {volume} {95}},\ \bibinfo
		{pages} {033208} (\bibinfo {year} {2017})}\BibitemShut {NoStop}%
	\bibitem [{\citenamefont {Borghesi}\ \emph {et~al.}(1998)\citenamefont
		{Borghesi}, \citenamefont {MacKinnon}, \citenamefont {Bell}, \citenamefont
		{Gaillard},\ and\ \citenamefont {Willi}}]{Borghesi98}%
	\BibitemOpen
	\bibfield  {author} {\bibinfo {author} {\bibfnamefont {M.}~\bibnamefont
			{Borghesi}}, \bibinfo {author} {\bibfnamefont {A.}~\bibnamefont {MacKinnon}},
		\bibinfo {author} {\bibfnamefont {A.}~\bibnamefont {Bell}}, \bibinfo {author}
		{\bibfnamefont {R.}~\bibnamefont {Gaillard}}, \ and\ \bibinfo {author}
		{\bibfnamefont {O.}~\bibnamefont {Willi}},\ }\bibfield  {title} {\enquote
		{\bibinfo {title} {Megagauss magnetic field generation and plasma jet
				formation on solid targets irradiated by an ultraintense picosecond laser
				pulse},}\ }\href@noop {} {\bibfield  {journal} {\bibinfo  {journal} {Phys.
				Rev. Lett.}\ }\textbf {\bibinfo {volume} {81}},\ \bibinfo {pages} {112}
		(\bibinfo {year} {1998})}\BibitemShut {NoStop}%
	\bibitem [{\citenamefont {Tatarakis}\ \emph
		{et~al.}(2002{\natexlab{a}})\citenamefont {Tatarakis}, \citenamefont {Gopal},
		\citenamefont {Watts}, \citenamefont {Beg}, \citenamefont {Dangor},
		\citenamefont {Krushelnick}, \citenamefont {Wagner}, \citenamefont {Norreys},
		\citenamefont {Clark}, \citenamefont {Zepf} \emph {et~al.}}]{Tatarakis02}%
	\BibitemOpen
	\bibfield  {author} {\bibinfo {author} {\bibfnamefont {M.}~\bibnamefont
			{Tatarakis}}, \bibinfo {author} {\bibfnamefont {A.}~\bibnamefont {Gopal}},
		\bibinfo {author} {\bibfnamefont {I.}~\bibnamefont {Watts}}, \bibinfo
		{author} {\bibfnamefont {F.}~\bibnamefont {Beg}}, \bibinfo {author}
		{\bibfnamefont {A.}~\bibnamefont {Dangor}}, \bibinfo {author} {\bibfnamefont
			{K.}~\bibnamefont {Krushelnick}}, \bibinfo {author} {\bibfnamefont
			{U.}~\bibnamefont {Wagner}}, \bibinfo {author} {\bibfnamefont
			{P.}~\bibnamefont {Norreys}}, \bibinfo {author} {\bibfnamefont
			{E.}~\bibnamefont {Clark}}, \bibinfo {author} {\bibfnamefont
			{M.}~\bibnamefont {Zepf}},  \emph {et~al.},\ }\bibfield  {title} {\enquote
		{\bibinfo {title} {Measurements of ultrastrong magnetic fields during
				relativistic laser--plasma interactions},}\ }\href@noop {} {\bibfield
		{journal} {\bibinfo  {journal} {Phys. Plasmas}\ }\textbf {\bibinfo {volume}
			{9}},\ \bibinfo {pages} {2244--2250} (\bibinfo {year}
		{2002}{\natexlab{a}})}\BibitemShut {NoStop}%
	\bibitem [{\citenamefont {Tatarakis}\ \emph
		{et~al.}(2002{\natexlab{b}})\citenamefont {Tatarakis}, \citenamefont {Watts},
		\citenamefont {Beg}, \citenamefont {Clark}, \citenamefont {Dangor},
		\citenamefont {Gopal}, \citenamefont {Haines}, \citenamefont {Norreys},
		\citenamefont {Wagner}, \citenamefont {Wei} \emph
		{et~al.}}]{Tatarakis02Nature}%
	\BibitemOpen
	\bibfield  {author} {\bibinfo {author} {\bibfnamefont {M.}~\bibnamefont
			{Tatarakis}}, \bibinfo {author} {\bibfnamefont {I.}~\bibnamefont {Watts}},
		\bibinfo {author} {\bibfnamefont {F.}~\bibnamefont {Beg}}, \bibinfo {author}
		{\bibfnamefont {E.}~\bibnamefont {Clark}}, \bibinfo {author} {\bibfnamefont
			{A.}~\bibnamefont {Dangor}}, \bibinfo {author} {\bibfnamefont
			{A.}~\bibnamefont {Gopal}}, \bibinfo {author} {\bibfnamefont
			{M.}~\bibnamefont {Haines}}, \bibinfo {author} {\bibfnamefont
			{P.}~\bibnamefont {Norreys}}, \bibinfo {author} {\bibfnamefont
			{U.}~\bibnamefont {Wagner}}, \bibinfo {author} {\bibfnamefont {M.-S.}\
			\bibnamefont {Wei}},  \emph {et~al.},\ }\bibfield  {title} {\enquote
		{\bibinfo {title} {Laser technology: Measuring huge magnetic fields},}\
	}\href@noop {} {\bibfield  {journal} {\bibinfo  {journal} {Nature}\ }\textbf
		{\bibinfo {volume} {415}},\ \bibinfo {pages} {280--280} (\bibinfo {year}
		{2002}{\natexlab{b}})}\BibitemShut {NoStop}%
	\bibitem [{\citenamefont {Wagner}\ \emph {et~al.}(2004)\citenamefont {Wagner},
		\citenamefont {Tatarakis}, \citenamefont {Gopal}, \citenamefont {Beg},
		\citenamefont {Clark}, \citenamefont {Dangor}, \citenamefont {Evans},
		\citenamefont {Haines}, \citenamefont {Mangles}, \citenamefont {Norreys}
		\emph {et~al.}}]{Wagner04}%
	\BibitemOpen
	\bibfield  {author} {\bibinfo {author} {\bibfnamefont {U.}~\bibnamefont
			{Wagner}}, \bibinfo {author} {\bibfnamefont {M.}~\bibnamefont {Tatarakis}},
		\bibinfo {author} {\bibfnamefont {A.}~\bibnamefont {Gopal}}, \bibinfo
		{author} {\bibfnamefont {F.}~\bibnamefont {Beg}}, \bibinfo {author}
		{\bibfnamefont {E.}~\bibnamefont {Clark}}, \bibinfo {author} {\bibfnamefont
			{A.}~\bibnamefont {Dangor}}, \bibinfo {author} {\bibfnamefont
			{R.}~\bibnamefont {Evans}}, \bibinfo {author} {\bibfnamefont
			{M.}~\bibnamefont {Haines}}, \bibinfo {author} {\bibfnamefont
			{S.}~\bibnamefont {Mangles}}, \bibinfo {author} {\bibfnamefont
			{P.}~\bibnamefont {Norreys}},  \emph {et~al.},\ }\bibfield  {title} {\enquote
		{\bibinfo {title} {Laboratory measurements of 0.7 gg magnetic fields
				generated during high-intensity laser interactions with dense plasmas},}\
	}\href@noop {} {\bibfield  {journal} {\bibinfo  {journal} {Phys. Rev. E}\
		}\textbf {\bibinfo {volume} {70}},\ \bibinfo {pages} {026401} (\bibinfo
		{year} {2004})}\BibitemShut {NoStop}%
	\bibitem [{\citenamefont {Manuel}\ \emph {et~al.}(2012)\citenamefont {Manuel},
		\citenamefont {Li}, \citenamefont {S\'eguin}, \citenamefont {Frenje},
		\citenamefont {Casey}, \citenamefont {Petrasso}, \citenamefont {Hu},
		\citenamefont {Betti}, \citenamefont {Hager}, \citenamefont {Meyerhofer},\
		and\ \citenamefont {Smalyuk}}]{Manuel12}%
	\BibitemOpen
	\bibfield  {author} {\bibinfo {author} {\bibfnamefont {M.~J.-E.}\
			\bibnamefont {Manuel}}, \bibinfo {author} {\bibfnamefont {C.~K.}\
			\bibnamefont {Li}}, \bibinfo {author} {\bibfnamefont {F.~H.}\ \bibnamefont
			{S\'eguin}}, \bibinfo {author} {\bibfnamefont {J.}~\bibnamefont {Frenje}},
		\bibinfo {author} {\bibfnamefont {D.~T.}\ \bibnamefont {Casey}}, \bibinfo
		{author} {\bibfnamefont {R.~D.}\ \bibnamefont {Petrasso}}, \bibinfo {author}
		{\bibfnamefont {S.~X.}\ \bibnamefont {Hu}}, \bibinfo {author} {\bibfnamefont
			{R.}~\bibnamefont {Betti}}, \bibinfo {author} {\bibfnamefont {J.~D.}\
			\bibnamefont {Hager}}, \bibinfo {author} {\bibfnamefont {D.~D.}\ \bibnamefont
			{Meyerhofer}}, \ and\ \bibinfo {author} {\bibfnamefont {V.~A.}\ \bibnamefont
			{Smalyuk}},\ }\bibfield  {title} {\enquote {\bibinfo {title} {First
				measurements of {R}ayleigh-{T}ylor-induced magnetic fields in laser-produced
				plasmas},}\ }\href {\doibase 10.1103/PhysRevLett.108.255006} {\bibfield
		{journal} {\bibinfo  {journal} {Phys. Rev. Lett.}\ }\textbf {\bibinfo
			{volume} {108}},\ \bibinfo {pages} {255006} (\bibinfo {year}
		{2012})}\BibitemShut {NoStop}%
	\bibitem [{\citenamefont {Gao}\ \emph {et~al.}(2012)\citenamefont {Gao},
		\citenamefont {Nilson}, \citenamefont {Igumenschev}, \citenamefont {Hu},
		\citenamefont {Davies}, \citenamefont {Stoeckl}, \citenamefont {Haines},
		\citenamefont {Froula}, \citenamefont {Betti},\ and\ \citenamefont
		{Meyerhofer}}]{Gao12}%
	\BibitemOpen
	\bibfield  {author} {\bibinfo {author} {\bibfnamefont {L.}~\bibnamefont
			{Gao}}, \bibinfo {author} {\bibfnamefont {P.}~\bibnamefont {Nilson}},
		\bibinfo {author} {\bibfnamefont {I.}~\bibnamefont {Igumenschev}}, \bibinfo
		{author} {\bibfnamefont {S.}~\bibnamefont {Hu}}, \bibinfo {author}
		{\bibfnamefont {J.}~\bibnamefont {Davies}}, \bibinfo {author} {\bibfnamefont
			{C.}~\bibnamefont {Stoeckl}}, \bibinfo {author} {\bibfnamefont
			{M.}~\bibnamefont {Haines}}, \bibinfo {author} {\bibfnamefont
			{D.}~\bibnamefont {Froula}}, \bibinfo {author} {\bibfnamefont
			{R.}~\bibnamefont {Betti}}, \ and\ \bibinfo {author} {\bibfnamefont
			{D.}~\bibnamefont {Meyerhofer}},\ }\bibfield  {title} {\enquote {\bibinfo
			{title} {Magnetic field generation by the {R}ayleigh-{T}aylor instability in
				laser-driven planar plastic targets},}\ }\href@noop {} {\bibfield  {journal}
		{\bibinfo  {journal} {Phys. Rev. Lett.}\ }\textbf {\bibinfo {volume} {109}},\
		\bibinfo {pages} {115001} (\bibinfo {year} {2012})}\BibitemShut {NoStop}%
	\bibitem [{\citenamefont {Gao}\ \emph {et~al.}(2015)\citenamefont {Gao},
		\citenamefont {Nilson}, \citenamefont {Igumenshchev}, \citenamefont {Haines},
		\citenamefont {Froula}, \citenamefont {Betti},\ and\ \citenamefont
		{Meyerhofer}}]{Gao15}%
	\BibitemOpen
	\bibfield  {author} {\bibinfo {author} {\bibfnamefont {L.}~\bibnamefont
			{Gao}}, \bibinfo {author} {\bibfnamefont {P.}~\bibnamefont {Nilson}},
		\bibinfo {author} {\bibfnamefont {I.}~\bibnamefont {Igumenshchev}}, \bibinfo
		{author} {\bibfnamefont {M.}~\bibnamefont {Haines}}, \bibinfo {author}
		{\bibfnamefont {D.}~\bibnamefont {Froula}}, \bibinfo {author} {\bibfnamefont
			{R.}~\bibnamefont {Betti}}, \ and\ \bibinfo {author} {\bibfnamefont
			{D.}~\bibnamefont {Meyerhofer}},\ }\bibfield  {title} {\enquote {\bibinfo
			{title} {Precision mapping of laser-driven magnetic fields and their
				evolution in high-energy-density plasmas},}\ }\href@noop {} {\bibfield
		{journal} {\bibinfo  {journal} {Phys. Rev. Lett.}\ }\textbf {\bibinfo
			{volume} {114}},\ \bibinfo {pages} {215003} (\bibinfo {year}
		{2015})}\BibitemShut {NoStop}%
	\bibitem [{\citenamefont {Sheffield}\ \emph {et~al.}(2010)\citenamefont
		{Sheffield}, \citenamefont {Froula}, \citenamefont {Glenzer},\ and\
		\citenamefont {Luhmann~Jr}}]{Sheffield10}%
	\BibitemOpen
	\bibfield  {author} {\bibinfo {author} {\bibfnamefont {J.}~\bibnamefont
			{Sheffield}}, \bibinfo {author} {\bibfnamefont {D.}~\bibnamefont {Froula}},
		\bibinfo {author} {\bibfnamefont {S.~H.}\ \bibnamefont {Glenzer}}, \ and\
		\bibinfo {author} {\bibfnamefont {N.~C.}\ \bibnamefont {Luhmann~Jr}},\
	}\href@noop {} {\emph {\bibinfo {title} {Plasma scattering of electromagnetic
				radiation: theory and measurement techniques}}}\ (\bibinfo  {publisher}
	{Academic press},\ \bibinfo {year} {2010})\BibitemShut {NoStop}%
	\bibitem [{\citenamefont {Stix}(1992)}]{Stix92}%
	\BibitemOpen
	\bibfield  {author} {\bibinfo {author} {\bibfnamefont {T.}~\bibnamefont
			{Stix}},\ }\href@noop {} {\emph {\bibinfo {title} {Waves in Plasmas}}}\
	(\bibinfo  {publisher} {American Inst. of Physics},\ \bibinfo {year}
	{1992})\BibitemShut {NoStop}%
	\bibitem [{\citenamefont {Drake}\ \emph {et~al.}(1974)\citenamefont {Drake},
		\citenamefont {Kaw}, \citenamefont {Lee}, \citenamefont {Schmid},
		\citenamefont {Liu},\ and\ \citenamefont {Rosenbluth}}]{Drake74}%
	\BibitemOpen
	\bibfield  {author} {\bibinfo {author} {\bibfnamefont {J.~F.}\ \bibnamefont
			{Drake}}, \bibinfo {author} {\bibfnamefont {P.~K.}\ \bibnamefont {Kaw}},
		\bibinfo {author} {\bibfnamefont {Y.-C.}\ \bibnamefont {Lee}}, \bibinfo
		{author} {\bibfnamefont {G.}~\bibnamefont {Schmid}}, \bibinfo {author}
		{\bibfnamefont {C.~S.}\ \bibnamefont {Liu}}, \ and\ \bibinfo {author}
		{\bibfnamefont {M.~N.}\ \bibnamefont {Rosenbluth}},\ }\bibfield  {title}
	{\enquote {\bibinfo {title} {Parametric instabilities of electromagnetic
				waves in plasmas},}\ }\href@noop {} {\bibfield  {journal} {\bibinfo
			{journal} {Phys. Fluids}\ }\textbf {\bibinfo {volume} {17}},\ \bibinfo
		{pages} {778--785} (\bibinfo {year} {1974})}\BibitemShut {NoStop}%
	\bibitem [{\citenamefont {Shi}, \citenamefont {Qin},\ and\ \citenamefont
		{Fisch}(2017{\natexlab{a}})}]{Shi17scatter}%
	\BibitemOpen
	\bibfield  {author} {\bibinfo {author} {\bibfnamefont {Y.}~\bibnamefont
			{Shi}}, \bibinfo {author} {\bibfnamefont {H.}~\bibnamefont {Qin}}, \ and\
		\bibinfo {author} {\bibfnamefont {N.~J.}\ \bibnamefont {Fisch}},\ }\bibfield
	{title} {\enquote {\bibinfo {title} {Three-wave scattering in magnetized
				plasmas: From cold fluid to quantized lagrangian},}\ }\href {\doibase
		10.1103/PhysRevE.96.023204} {\bibfield  {journal} {\bibinfo  {journal} {Phys.
				Rev. E}\ }\textbf {\bibinfo {volume} {96}},\ \bibinfo {pages} {023204}
		(\bibinfo {year} {2017}{\natexlab{a}})}\BibitemShut {NoStop}%
	\bibitem [{\citenamefont {Wang}\ \emph {et~al.}(2015)\citenamefont {Wang},
		\citenamefont {Gibbon}, \citenamefont {Sheng},\ and\ \citenamefont
		{Li}}]{Wang15}%
	\BibitemOpen
	\bibfield  {author} {\bibinfo {author} {\bibfnamefont {W.-M.}\ \bibnamefont
			{Wang}}, \bibinfo {author} {\bibfnamefont {P.}~\bibnamefont {Gibbon}},
		\bibinfo {author} {\bibfnamefont {Z.-M.}\ \bibnamefont {Sheng}}, \ and\
		\bibinfo {author} {\bibfnamefont {Y.-T.}\ \bibnamefont {Li}},\ }\bibfield
	{title} {\enquote {\bibinfo {title} {Magnetically assisted fast ignition},}\
	}\href {\doibase 10.1103/PhysRevLett.114.015001} {\bibfield  {journal}
		{\bibinfo  {journal} {Phys. Rev. Lett.}\ }\textbf {\bibinfo {volume} {114}},\
		\bibinfo {pages} {015001} (\bibinfo {year} {2015})}\BibitemShut {NoStop}%
	\bibitem [{\citenamefont {Barnak}\ \emph {et~al.}(2017)\citenamefont {Barnak},
		\citenamefont {Davies}, \citenamefont {Betti}, \citenamefont {Bonino},
		\citenamefont {Campbell}, \citenamefont {Glebov}, \citenamefont {Harding},
		\citenamefont {Knauer}, \citenamefont {Regan}, \citenamefont {Sefkow} \emph
		{et~al.}}]{Barnak17}%
	\BibitemOpen
	\bibfield  {author} {\bibinfo {author} {\bibfnamefont {D.}~\bibnamefont
			{Barnak}}, \bibinfo {author} {\bibfnamefont {J.}~\bibnamefont {Davies}},
		\bibinfo {author} {\bibfnamefont {R.}~\bibnamefont {Betti}}, \bibinfo
		{author} {\bibfnamefont {M.}~\bibnamefont {Bonino}}, \bibinfo {author}
		{\bibfnamefont {E.}~\bibnamefont {Campbell}}, \bibinfo {author}
		{\bibfnamefont {V.~Y.}\ \bibnamefont {Glebov}}, \bibinfo {author}
		{\bibfnamefont {D.}~\bibnamefont {Harding}}, \bibinfo {author} {\bibfnamefont
			{J.}~\bibnamefont {Knauer}}, \bibinfo {author} {\bibfnamefont
			{S.}~\bibnamefont {Regan}}, \bibinfo {author} {\bibfnamefont
			{A.}~\bibnamefont {Sefkow}},  \emph {et~al.},\ }\bibfield  {title} {\enquote
		{\bibinfo {title} {Laser-driven magnetized liner inertial fusion on omega},}\
	}\href@noop {} {\bibfield  {journal} {\bibinfo  {journal} {Phys. Plasmas}\
		}\textbf {\bibinfo {volume} {24}},\ \bibinfo {pages} {056310} (\bibinfo
		{year} {2017})}\BibitemShut {NoStop}%
	\bibitem [{\citenamefont {Milroy}, \citenamefont {Capjack},\ and\ \citenamefont
		{James}(1979)}]{Milroy79}%
	\BibitemOpen
	\bibfield  {author} {\bibinfo {author} {\bibfnamefont {R.}~\bibnamefont
			{Milroy}}, \bibinfo {author} {\bibfnamefont {C.}~\bibnamefont {Capjack}}, \
		and\ \bibinfo {author} {\bibfnamefont {C.}~\bibnamefont {James}},\ }\bibfield
	{title} {\enquote {\bibinfo {title} {Plasma laser pulse amplifier using
				induced {R}aman or {B}rillouin processes},}\ }\href@noop {} {\bibfield
		{journal} {\bibinfo  {journal} {Phys. Fluids}\ }\textbf {\bibinfo {volume}
			{22}},\ \bibinfo {pages} {1922--1931} (\bibinfo {year} {1979})}\BibitemShut
	{NoStop}%
	\bibitem [{\citenamefont {Malkin}, \citenamefont {Shvets},\ and\ \citenamefont
		{Fisch}(1999)}]{Malkin99}%
	\BibitemOpen
	\bibfield  {author} {\bibinfo {author} {\bibfnamefont {V.}~\bibnamefont
			{Malkin}}, \bibinfo {author} {\bibfnamefont {G.}~\bibnamefont {Shvets}}, \
		and\ \bibinfo {author} {\bibfnamefont {N.}~\bibnamefont {Fisch}},\ }\bibfield
	{title} {\enquote {\bibinfo {title} {Fast compression of laser beams to
				highly overcritical powers},}\ }\href@noop {} {\bibfield  {journal} {\bibinfo
			{journal} {Phys. Rev. Lett.}\ }\textbf {\bibinfo {volume} {82}},\ \bibinfo
		{pages} {4448} (\bibinfo {year} {1999})}\BibitemShut {NoStop}%
	\bibitem [{\citenamefont {Malkin}, \citenamefont {Fisch},\ and\ \citenamefont
		{Wurtele}(2007)}]{Malkin07}%
	\BibitemOpen
	\bibfield  {author} {\bibinfo {author} {\bibfnamefont {V.}~\bibnamefont
			{Malkin}}, \bibinfo {author} {\bibfnamefont {N.}~\bibnamefont {Fisch}}, \
		and\ \bibinfo {author} {\bibfnamefont {J.}~\bibnamefont {Wurtele}},\
	}\bibfield  {title} {\enquote {\bibinfo {title} {Compression of powerful
				x-ray pulses to attosecond durations by stimulated {R}aman backscattering in
				plasmas},}\ }\href@noop {} {\bibfield  {journal} {\bibinfo  {journal} {Phys.
				Rev. E}\ }\textbf {\bibinfo {volume} {75}},\ \bibinfo {pages} {026404}
		(\bibinfo {year} {2007})}\BibitemShut {NoStop}%
	\bibitem [{\citenamefont {Andreev}\ \emph {et~al.}(2006)\citenamefont
		{Andreev}, \citenamefont {Riconda}, \citenamefont {Tikhonchuk},\ and\
		\citenamefont {Weber}}]{Andreev06}%
	\BibitemOpen
	\bibfield  {author} {\bibinfo {author} {\bibfnamefont {A.}~\bibnamefont
			{Andreev}}, \bibinfo {author} {\bibfnamefont {C.}~\bibnamefont {Riconda}},
		\bibinfo {author} {\bibfnamefont {V.}~\bibnamefont {Tikhonchuk}}, \ and\
		\bibinfo {author} {\bibfnamefont {S.}~\bibnamefont {Weber}},\ }\bibfield
	{title} {\enquote {\bibinfo {title} {Short light pulse amplification and
				compression by stimulated {B}rillouin scattering in plasmas in the strong
				coupling regime},}\ }\href@noop {} {\bibfield  {journal} {\bibinfo  {journal}
			{Phys. Plasmas}\ }\textbf {\bibinfo {volume} {13}},\ \bibinfo {pages}
		{053110} (\bibinfo {year} {2006})}\BibitemShut {NoStop}%
	\bibitem [{\citenamefont {Edwards}\ \emph {et~al.}(2016)\citenamefont
		{Edwards}, \citenamefont {Jia}, \citenamefont {Mikhailova},\ and\
		\citenamefont {Fisch}}]{Edwards16}%
	\BibitemOpen
	\bibfield  {author} {\bibinfo {author} {\bibfnamefont {M.~R.}\ \bibnamefont
			{Edwards}}, \bibinfo {author} {\bibfnamefont {Q.}~\bibnamefont {Jia}},
		\bibinfo {author} {\bibfnamefont {J.~M.}\ \bibnamefont {Mikhailova}}, \ and\
		\bibinfo {author} {\bibfnamefont {N.~J.}\ \bibnamefont {Fisch}},\ }\bibfield
	{title} {\enquote {\bibinfo {title} {Short-pulse amplification by strongly
				coupled stimulated {B}rillouin scattering},}\ }\href@noop {} {\bibfield
		{journal} {\bibinfo  {journal} {Phys. Plasmas}\ }\textbf {\bibinfo {volume}
			{23}},\ \bibinfo {pages} {083122} (\bibinfo {year} {2016})}\BibitemShut
	{NoStop}%
	\bibitem [{\citenamefont {Edwards}, \citenamefont {Mikhailova},\ and\
		\citenamefont {Fisch}(2017)}]{Edwards17}%
	\BibitemOpen
	\bibfield  {author} {\bibinfo {author} {\bibfnamefont {M.~R.}\ \bibnamefont
			{Edwards}}, \bibinfo {author} {\bibfnamefont {J.~M.}\ \bibnamefont
			{Mikhailova}}, \ and\ \bibinfo {author} {\bibfnamefont {N.~J.}\ \bibnamefont
			{Fisch}},\ }\bibfield  {title} {\enquote {\bibinfo {title} {X-ray
				amplification by stimulated {B}rillouin scattering},}\ }\href {\doibase
		10.1103/PhysRevE.96.023209} {\bibfield  {journal} {\bibinfo  {journal} {Phys.
				Rev. E}\ }\textbf {\bibinfo {volume} {96}},\ \bibinfo {pages} {023209}
		(\bibinfo {year} {2017})}\BibitemShut {NoStop}%
	\bibitem [{\citenamefont {Maine}\ \emph {et~al.}(1988)\citenamefont {Maine},
		\citenamefont {Strickland}, \citenamefont {Bado}, \citenamefont {Pessot},\
		and\ \citenamefont {Mourou}}]{Maine88}%
	\BibitemOpen
	\bibfield  {author} {\bibinfo {author} {\bibfnamefont {P.}~\bibnamefont
			{Maine}}, \bibinfo {author} {\bibfnamefont {D.}~\bibnamefont {Strickland}},
		\bibinfo {author} {\bibfnamefont {P.}~\bibnamefont {Bado}}, \bibinfo {author}
		{\bibfnamefont {M.}~\bibnamefont {Pessot}}, \ and\ \bibinfo {author}
		{\bibfnamefont {G.}~\bibnamefont {Mourou}},\ }\bibfield  {title} {\enquote
		{\bibinfo {title} {Generation of ultrahigh peak power pulses by chirped pulse
				amplification},}\ }\href@noop {} {\bibfield  {journal} {\bibinfo  {journal}
			{IEEE J. Quantum Elect.}\ }\textbf {\bibinfo {volume} {24}},\ \bibinfo
		{pages} {398--403} (\bibinfo {year} {1988})}\BibitemShut {NoStop}%
	\bibitem [{\citenamefont {Shi}, \citenamefont {Qin},\ and\ \citenamefont
		{Fisch}(2017{\natexlab{b}})}]{Shi17laser}%
	\BibitemOpen
	\bibfield  {author} {\bibinfo {author} {\bibfnamefont {Y.}~\bibnamefont
			{Shi}}, \bibinfo {author} {\bibfnamefont {H.}~\bibnamefont {Qin}}, \ and\
		\bibinfo {author} {\bibfnamefont {N.~J.}\ \bibnamefont {Fisch}},\ }\bibfield
	{title} {\enquote {\bibinfo {title} {Laser-pulse compression using magnetized
				plasmas},}\ }\href@noop {} {\bibfield  {journal} {\bibinfo  {journal} {Phys.
				Rev. E}\ }\textbf {\bibinfo {volume} {95}},\ \bibinfo {pages} {023211}
		(\bibinfo {year} {2017}{\natexlab{b}})}\BibitemShut {NoStop}%
	\bibitem [{\citenamefont {Shi}, \citenamefont {Fisch},\ and\ \citenamefont
		{Qin}(2016)}]{Shi16QED}%
	\BibitemOpen
	\bibfield  {author} {\bibinfo {author} {\bibfnamefont {Y.}~\bibnamefont
			{Shi}}, \bibinfo {author} {\bibfnamefont {N.~J.}\ \bibnamefont {Fisch}}, \
		and\ \bibinfo {author} {\bibfnamefont {H.}~\bibnamefont {Qin}},\ }\bibfield
	{title} {\enquote {\bibinfo {title} {Effective-action approach to wave
				propagation in scalar qed plasmas},}\ }\href@noop {} {\bibfield  {journal}
		{\bibinfo  {journal} {Phys. Rev. A}\ }\textbf {\bibinfo {volume} {94}},\
		\bibinfo {pages} {012124} (\bibinfo {year} {2016})}\BibitemShut {NoStop}%
	\bibitem [{\citenamefont {Kaup}, \citenamefont {Reiman},\ and\ \citenamefont
		{Bers}(1979)}]{Kaup79}%
	\BibitemOpen
	\bibfield  {author} {\bibinfo {author} {\bibfnamefont {D.~J.}\ \bibnamefont
			{Kaup}}, \bibinfo {author} {\bibfnamefont {A.}~\bibnamefont {Reiman}}, \ and\
		\bibinfo {author} {\bibfnamefont {A.}~\bibnamefont {Bers}},\ }\bibfield
	{title} {\enquote {\bibinfo {title} {Space-time evolution of nonlinear
				three-wave interactions. i. interaction in a homogeneous medium},}\
	}\href@noop {} {\bibfield  {journal} {\bibinfo  {journal} {Rev. Mod. Phys.}\
		}\textbf {\bibinfo {volume} {51}},\ \bibinfo {pages} {275--309} (\bibinfo
		{year} {1979})}\BibitemShut {NoStop}%
	\bibitem [{\citenamefont {Sj{\"o}lund}\ and\ \citenamefont
		{Stenflo}(1967)}]{Sjolund67}%
	\BibitemOpen
	\bibfield  {author} {\bibinfo {author} {\bibfnamefont {A.}~\bibnamefont
			{Sj{\"o}lund}}\ and\ \bibinfo {author} {\bibfnamefont {L.}~\bibnamefont
			{Stenflo}},\ }\bibfield  {title} {\enquote {\bibinfo {title} {Non-linear
				coupling in a magnetized plasma},}\ }\href {\doibase 10.1007/BF01326195}
	{\bibfield  {journal} {\bibinfo  {journal} {Z. Phys. A-Hadron. Nucl.}\
		}\textbf {\bibinfo {volume} {204}},\ \bibinfo {pages} {211--214} (\bibinfo
		{year} {1967})}\BibitemShut {NoStop}%
	\bibitem [{\citenamefont {Galloway}\ and\ \citenamefont
		{Kim}(1971)}]{Galloway71}%
	\BibitemOpen
	\bibfield  {author} {\bibinfo {author} {\bibfnamefont {J.~J.}\ \bibnamefont
			{Galloway}}\ and\ \bibinfo {author} {\bibfnamefont {H.}~\bibnamefont {Kim}},\
	}\bibfield  {title} {\enquote {\bibinfo {title} {Lagrangian approach to
				non-linear wave interactions in a warm plasma},}\ }\href@noop {} {\bibfield
		{journal} {\bibinfo  {journal} {J. Plasma Phys.}\ }\textbf {\bibinfo {volume}
			{6}},\ \bibinfo {pages} {53--72} (\bibinfo {year} {1971})}\BibitemShut
	{NoStop}%
	\bibitem [{\citenamefont {Boyd}\ and\ \citenamefont {Turner}(1978)}]{Boyd78}%
	\BibitemOpen
	\bibfield  {author} {\bibinfo {author} {\bibfnamefont {T.~J.~M.}\
			\bibnamefont {Boyd}}\ and\ \bibinfo {author} {\bibfnamefont {J.~G.}\
			\bibnamefont {Turner}},\ }\bibfield  {title} {\enquote {\bibinfo {title}
			{Three and four wave interactions in plasmas},}\ }\href@noop {} {\bibfield
		{journal} {\bibinfo  {journal} {J. Math. Phys.}\ }\textbf {\bibinfo {volume}
			{19}},\ \bibinfo {pages} {1403--1413} (\bibinfo {year} {1978})}\BibitemShut
	{NoStop}%
	\bibitem [{\citenamefont {Shivamoggi}(1982)}]{Shivamoggi82}%
	\BibitemOpen
	\bibfield  {author} {\bibinfo {author} {\bibfnamefont {B.~K.}\ \bibnamefont
			{Shivamoggi}},\ }\bibfield  {title} {\enquote {\bibinfo {title} {Kinetic
				theory of three-wave interaction in a magnetised, inhomogeneous plasma},}\
	}\href@noop {} {\bibfield  {journal} {\bibinfo  {journal} {Phys. Scripta}\
		}\textbf {\bibinfo {volume} {25}},\ \bibinfo {pages} {637} (\bibinfo {year}
		{1982})}\BibitemShut {NoStop}%
	\bibitem [{\citenamefont {Ram}(1982)}]{Ram82}%
	\BibitemOpen
	\bibfield  {author} {\bibinfo {author} {\bibfnamefont {S.}~\bibnamefont
			{Ram}},\ }\bibfield  {title} {\enquote {\bibinfo {title} {Nonlinear
				scattering from electron bernstein modes in a plasma},}\ }\href@noop {}
	{\bibfield  {journal} {\bibinfo  {journal} {Plasma Physics}\ }\textbf
		{\bibinfo {volume} {24}},\ \bibinfo {pages} {885} (\bibinfo {year}
		{1982})}\BibitemShut {NoStop}%
	\bibitem [{\citenamefont {Boyd}\ and\ \citenamefont {Rankin}(1985)}]{Boyd85}%
	\BibitemOpen
	\bibfield  {author} {\bibinfo {author} {\bibfnamefont {T.~J.~M.}\
			\bibnamefont {Boyd}}\ and\ \bibinfo {author} {\bibfnamefont {R.}~\bibnamefont
			{Rankin}},\ }\bibfield  {title} {\enquote {\bibinfo {title} {Kinetic theory
				of stimulated {R}aman scattering from a magnetized plasma},}\ }\href@noop {}
	{\bibfield  {journal} {\bibinfo  {journal} {J. Plasma Phys.}\ }\textbf
		{\bibinfo {volume} {33}},\ \bibinfo {pages} {303--319} (\bibinfo {year}
		{1985})}\BibitemShut {NoStop}%
	\bibitem [{\citenamefont {Cohen}(1987)}]{Cohen87}%
	\BibitemOpen
	\bibfield  {author} {\bibinfo {author} {\bibfnamefont {B.~I.}\ \bibnamefont
			{Cohen}},\ }\bibfield  {title} {\enquote {\bibinfo {title} {Compact
				dispersion relations for parametric instabilities of electromagnetic waves in
				magnetized plasmas},}\ }\href@noop {} {\bibfield  {journal} {\bibinfo
			{journal} {Phys. Fluids}\ }\textbf {\bibinfo {volume} {30}},\ \bibinfo
		{pages} {2676--2680} (\bibinfo {year} {1987})}\BibitemShut {NoStop}%
	\bibitem [{\citenamefont {Stefan}, \citenamefont {Krall},\ and\ \citenamefont
		{McBride}(1987)}]{Stefan87}%
	\BibitemOpen
	\bibfield  {author} {\bibinfo {author} {\bibfnamefont {V.}~\bibnamefont
			{Stefan}}, \bibinfo {author} {\bibfnamefont {N.}~\bibnamefont {Krall}}, \
		and\ \bibinfo {author} {\bibfnamefont {J.}~\bibnamefont {McBride}},\
	}\bibfield  {title} {\enquote {\bibinfo {title} {The nonlinear eikonal
				relation of a weakly inhomogeneous magnetized plasma upon the action of
				arbitrarily polarized finite wavelength electromagnetic waves},}\ }\href@noop
	{} {\bibfield  {journal} {\bibinfo  {journal} {Phys. Fluids}\ }\textbf
		{\bibinfo {volume} {30}},\ \bibinfo {pages} {3703--3712} (\bibinfo {year}
		{1987})}\BibitemShut {NoStop}%
	\bibitem [{\citenamefont {Laham}, \citenamefont {Nasser},\ and\ \citenamefont
		{Khateeb}(1998)}]{Laham98}%
	\BibitemOpen
	\bibfield  {author} {\bibinfo {author} {\bibfnamefont {N.~M.}\ \bibnamefont
			{Laham}}, \bibinfo {author} {\bibfnamefont {A.~S.~A.}\ \bibnamefont
			{Nasser}}, \ and\ \bibinfo {author} {\bibfnamefont {A.~M.}\ \bibnamefont
			{Khateeb}},\ }\bibfield  {title} {\enquote {\bibinfo {title} {Effects of
				axial magnetic fields on backward {R}aman scattering in inhomogeneous
				plasmas},}\ }\href@noop {} {\bibfield  {journal} {\bibinfo  {journal} {Phys.
				Scripta}\ }\textbf {\bibinfo {volume} {57}},\ \bibinfo {pages} {253}
		(\bibinfo {year} {1998})}\BibitemShut {NoStop}%
	\bibitem [{\citenamefont {Sanuki}\ and\ \citenamefont
		{Schmidt}(1977)}]{Sanuki77}%
	\BibitemOpen
	\bibfield  {author} {\bibinfo {author} {\bibfnamefont {H.}~\bibnamefont
			{Sanuki}}\ and\ \bibinfo {author} {\bibfnamefont {G.}~\bibnamefont
			{Schmidt}},\ }\bibfield  {title} {\enquote {\bibinfo {title} {Parametric
				instabilities in magnetized plasma},}\ }\href@noop {} {\bibfield  {journal}
		{\bibinfo  {journal} {J. Phys. Soc. Jpn.}\ }\textbf {\bibinfo {volume}
			{42}},\ \bibinfo {pages} {664--669} (\bibinfo {year} {1977})}\BibitemShut
	{NoStop}%
	\bibitem [{\citenamefont {Grebogi}\ and\ \citenamefont
		{Liu}(1980)}]{Grebogi80}%
	\BibitemOpen
	\bibfield  {author} {\bibinfo {author} {\bibfnamefont {C.}~\bibnamefont
			{Grebogi}}\ and\ \bibinfo {author} {\bibfnamefont {C.~S.}\ \bibnamefont
			{Liu}},\ }\bibfield  {title} {\enquote {\bibinfo {title} {{B}rillouin and
				{R}aman scattering of an extraordinary mode in a magnetized plasma},}\
	}\href@noop {} {\bibfield  {journal} {\bibinfo  {journal} {Phys. Fluids}\
		}\textbf {\bibinfo {volume} {23}},\ \bibinfo {pages} {1330--1335} (\bibinfo
		{year} {1980})}\BibitemShut {NoStop}%
	\bibitem [{\citenamefont {Dodin}\ and\ \citenamefont
		{Arefiev}(2017)}]{Dodin17}%
	\BibitemOpen
	\bibfield  {author} {\bibinfo {author} {\bibfnamefont {I.}~\bibnamefont
			{Dodin}}\ and\ \bibinfo {author} {\bibfnamefont {A.}~\bibnamefont
			{Arefiev}},\ }\bibfield  {title} {\enquote {\bibinfo {title} {Parametric
				decay of plasma waves near the upper-hybrid resonance},}\ }\href@noop {}
	{\bibfield  {journal} {\bibinfo  {journal} {Phys. Plasmas}\ }\textbf
		{\bibinfo {volume} {24}},\ \bibinfo {pages} {032119} (\bibinfo {year}
		{2017})}\BibitemShut {NoStop}%
	\bibitem [{\citenamefont {Stuart}\ \emph {et~al.}(1995)\citenamefont {Stuart},
		\citenamefont {Feit}, \citenamefont {Rubenchik}, \citenamefont {Shore},\ and\
		\citenamefont {Perry}}]{Stuart95}%
	\BibitemOpen
	\bibfield  {author} {\bibinfo {author} {\bibfnamefont {B.~C.}\ \bibnamefont
			{Stuart}}, \bibinfo {author} {\bibfnamefont {M.~D.}\ \bibnamefont {Feit}},
		\bibinfo {author} {\bibfnamefont {A.~M.}\ \bibnamefont {Rubenchik}}, \bibinfo
		{author} {\bibfnamefont {B.~W.}\ \bibnamefont {Shore}}, \ and\ \bibinfo
		{author} {\bibfnamefont {M.~D.}\ \bibnamefont {Perry}},\ }\bibfield  {title}
	{\enquote {\bibinfo {title} {Laser-induced damage in dielectrics with
				nanosecond to subpicosecond pulses},}\ }\href {\doibase
		10.1103/PhysRevLett.74.2248} {\bibfield  {journal} {\bibinfo  {journal}
			{Phys. Rev. Lett.}\ }\textbf {\bibinfo {volume} {74}},\ \bibinfo {pages}
		{2248--2251} (\bibinfo {year} {1995})}\BibitemShut {NoStop}%
	\bibitem [{\citenamefont {Canova}\ \emph {et~al.}(2007)\citenamefont {Canova},
		\citenamefont {Uteza}, \citenamefont {Chambaret}, \citenamefont {Flury},
		\citenamefont {Tonchev}, \citenamefont {Fechner},\ and\ \citenamefont
		{Parriaux}}]{Canova07}%
	\BibitemOpen
	\bibfield  {author} {\bibinfo {author} {\bibfnamefont {F.}~\bibnamefont
			{Canova}}, \bibinfo {author} {\bibfnamefont {O.}~\bibnamefont {Uteza}},
		\bibinfo {author} {\bibfnamefont {J.-P.}\ \bibnamefont {Chambaret}}, \bibinfo
		{author} {\bibfnamefont {M.}~\bibnamefont {Flury}}, \bibinfo {author}
		{\bibfnamefont {S.}~\bibnamefont {Tonchev}}, \bibinfo {author} {\bibfnamefont
			{R.}~\bibnamefont {Fechner}}, \ and\ \bibinfo {author} {\bibfnamefont
			{O.}~\bibnamefont {Parriaux}},\ }\bibfield  {title} {\enquote {\bibinfo
			{title} {High-efficiency, broad band, high-damage threshold high-index
				gratings for femtosecond pulse compression},}\ }\href@noop {} {\bibfield
		{journal} {\bibinfo  {journal} {Opt. Express}\ }\textbf {\bibinfo {volume}
			{15}},\ \bibinfo {pages} {15324--15334} (\bibinfo {year} {2007})}\BibitemShut
	{NoStop}%
	\bibitem [{\citenamefont {Obenschain}\ \emph {et~al.}(1996)\citenamefont
		{Obenschain}, \citenamefont {Bodner}, \citenamefont {Colombant},
		\citenamefont {Gerber}, \citenamefont {Lehmberg}, \citenamefont {McLean},
		\citenamefont {Mostovych}, \citenamefont {Pronko}, \citenamefont {Pawley},
		\citenamefont {Schmitt} \emph {et~al.}}]{Obenschain96}%
	\BibitemOpen
	\bibfield  {author} {\bibinfo {author} {\bibfnamefont {S.}~\bibnamefont
			{Obenschain}}, \bibinfo {author} {\bibfnamefont {S.}~\bibnamefont {Bodner}},
		\bibinfo {author} {\bibfnamefont {D.}~\bibnamefont {Colombant}}, \bibinfo
		{author} {\bibfnamefont {K.}~\bibnamefont {Gerber}}, \bibinfo {author}
		{\bibfnamefont {R.}~\bibnamefont {Lehmberg}}, \bibinfo {author}
		{\bibfnamefont {E.}~\bibnamefont {McLean}}, \bibinfo {author} {\bibfnamefont
			{A.}~\bibnamefont {Mostovych}}, \bibinfo {author} {\bibfnamefont
			{M.}~\bibnamefont {Pronko}}, \bibinfo {author} {\bibfnamefont
			{C.}~\bibnamefont {Pawley}}, \bibinfo {author} {\bibfnamefont
			{A.}~\bibnamefont {Schmitt}},  \emph {et~al.},\ }\bibfield  {title} {\enquote
		{\bibinfo {title} {The nike krf laser facility: Performance and initial
				target experiments},}\ }\href@noop {} {\bibfield  {journal} {\bibinfo
			{journal} {Phys. Plasmas}\ }\textbf {\bibinfo {volume} {3}},\ \bibinfo
		{pages} {2098--2107} (\bibinfo {year} {1996})}\BibitemShut {NoStop}%
	\bibitem [{\citenamefont {Emma}\ \emph {et~al.}(2010)\citenamefont {Emma},
		\citenamefont {Akre}, \citenamefont {Arthur}, \citenamefont {Bionta},
		\citenamefont {Bostedt}, \citenamefont {Bozek}, \citenamefont {Brachmann},
		\citenamefont {Bucksbaum}, \citenamefont {Coffee}, \citenamefont {Decker}
		\emph {et~al.}}]{Emma10}%
	\BibitemOpen
	\bibfield  {author} {\bibinfo {author} {\bibfnamefont {P.}~\bibnamefont
			{Emma}}, \bibinfo {author} {\bibfnamefont {R.}~\bibnamefont {Akre}}, \bibinfo
		{author} {\bibfnamefont {J.}~\bibnamefont {Arthur}}, \bibinfo {author}
		{\bibfnamefont {R.}~\bibnamefont {Bionta}}, \bibinfo {author} {\bibfnamefont
			{C.}~\bibnamefont {Bostedt}}, \bibinfo {author} {\bibfnamefont
			{J.}~\bibnamefont {Bozek}}, \bibinfo {author} {\bibfnamefont
			{A.}~\bibnamefont {Brachmann}}, \bibinfo {author} {\bibfnamefont
			{P.}~\bibnamefont {Bucksbaum}}, \bibinfo {author} {\bibfnamefont
			{R.}~\bibnamefont {Coffee}}, \bibinfo {author} {\bibfnamefont {F.-J.}\
			\bibnamefont {Decker}},  \emph {et~al.},\ }\bibfield  {title} {\enquote
		{\bibinfo {title} {First lasing and operation of an {\aa}ngstrom-wavelength
				free-electron laser},}\ }\href@noop {} {\bibfield  {journal} {\bibinfo
			{journal} {Nat. Photonics}\ }\textbf {\bibinfo {volume} {4}},\ \bibinfo
		{pages} {641--647} (\bibinfo {year} {2010})}\BibitemShut {NoStop}%
	\bibitem [{\citenamefont {Ishikawa}\ \emph {et~al.}(2012)\citenamefont
		{Ishikawa}, \citenamefont {Aoyagi}, \citenamefont {Asaka}, \citenamefont
		{Asano}, \citenamefont {Azumi}, \citenamefont {Bizen}, \citenamefont {Ego},
		\citenamefont {Fukami}, \citenamefont {Fukui}, \citenamefont {Furukawa} \emph
		{et~al.}}]{Ihikawa12}%
	\BibitemOpen
	\bibfield  {author} {\bibinfo {author} {\bibfnamefont {T.}~\bibnamefont
			{Ishikawa}}, \bibinfo {author} {\bibfnamefont {H.}~\bibnamefont {Aoyagi}},
		\bibinfo {author} {\bibfnamefont {T.}~\bibnamefont {Asaka}}, \bibinfo
		{author} {\bibfnamefont {Y.}~\bibnamefont {Asano}}, \bibinfo {author}
		{\bibfnamefont {N.}~\bibnamefont {Azumi}}, \bibinfo {author} {\bibfnamefont
			{T.}~\bibnamefont {Bizen}}, \bibinfo {author} {\bibfnamefont
			{H.}~\bibnamefont {Ego}}, \bibinfo {author} {\bibfnamefont {K.}~\bibnamefont
			{Fukami}}, \bibinfo {author} {\bibfnamefont {T.}~\bibnamefont {Fukui}},
		\bibinfo {author} {\bibfnamefont {Y.}~\bibnamefont {Furukawa}},  \emph
		{et~al.},\ }\bibfield  {title} {\enquote {\bibinfo {title} {A compact x-ray
				free-electron laser emitting in the sub-angstrom region},}\ }\href@noop {}
	{\bibfield  {journal} {\bibinfo  {journal} {Nat. Photonics}\ }\textbf
		{\bibinfo {volume} {6}},\ \bibinfo {pages} {540--544} (\bibinfo {year}
		{2012})}\BibitemShut {NoStop}%
	\bibitem [{\citenamefont {Glenzer}\ \emph {et~al.}(2011)\citenamefont
		{Glenzer}, \citenamefont {MacGowan}, \citenamefont {Meezan}, \citenamefont
		{Adams}, \citenamefont {Alfonso}, \citenamefont {Alger}, \citenamefont
		{Alherz}, \citenamefont {Alvarez}, \citenamefont {Alvarez}, \citenamefont
		{Amick} \emph {et~al.}}]{Glenzer11}%
	\BibitemOpen
	\bibfield  {author} {\bibinfo {author} {\bibfnamefont {S.}~\bibnamefont
			{Glenzer}}, \bibinfo {author} {\bibfnamefont {B.}~\bibnamefont {MacGowan}},
		\bibinfo {author} {\bibfnamefont {N.}~\bibnamefont {Meezan}}, \bibinfo
		{author} {\bibfnamefont {P.}~\bibnamefont {Adams}}, \bibinfo {author}
		{\bibfnamefont {J.}~\bibnamefont {Alfonso}}, \bibinfo {author} {\bibfnamefont
			{E.}~\bibnamefont {Alger}}, \bibinfo {author} {\bibfnamefont
			{Z.}~\bibnamefont {Alherz}}, \bibinfo {author} {\bibfnamefont
			{L.}~\bibnamefont {Alvarez}}, \bibinfo {author} {\bibfnamefont
			{S.}~\bibnamefont {Alvarez}}, \bibinfo {author} {\bibfnamefont
			{P.}~\bibnamefont {Amick}},  \emph {et~al.},\ }\bibfield  {title} {\enquote
		{\bibinfo {title} {Demonstration of ignition radiation temperatures in
				indirect-drive inertial confinement fusion hohlraums},}\ }\href@noop {}
	{\bibfield  {journal} {\bibinfo  {journal} {Phys. Rev. Lett.}\ }\textbf
		{\bibinfo {volume} {106}},\ \bibinfo {pages} {085004} (\bibinfo {year}
		{2011})}\BibitemShut {NoStop}%
	\bibitem [{\citenamefont {Craxton}\ \emph {et~al.}(2015)\citenamefont
		{Craxton}, \citenamefont {Anderson}, \citenamefont {Boehly}, \citenamefont
		{Goncharov}, \citenamefont {Harding}, \citenamefont {Knauer}, \citenamefont
		{McCrory}, \citenamefont {McKenty}, \citenamefont {Meyerhofer}, \citenamefont
		{Myatt} \emph {et~al.}}]{Craxton15}%
	\BibitemOpen
	\bibfield  {author} {\bibinfo {author} {\bibfnamefont {R.}~\bibnamefont
			{Craxton}}, \bibinfo {author} {\bibfnamefont {K.}~\bibnamefont {Anderson}},
		\bibinfo {author} {\bibfnamefont {T.}~\bibnamefont {Boehly}}, \bibinfo
		{author} {\bibfnamefont {V.}~\bibnamefont {Goncharov}}, \bibinfo {author}
		{\bibfnamefont {D.}~\bibnamefont {Harding}}, \bibinfo {author} {\bibfnamefont
			{J.}~\bibnamefont {Knauer}}, \bibinfo {author} {\bibfnamefont
			{R.}~\bibnamefont {McCrory}}, \bibinfo {author} {\bibfnamefont
			{P.}~\bibnamefont {McKenty}}, \bibinfo {author} {\bibfnamefont
			{D.}~\bibnamefont {Meyerhofer}}, \bibinfo {author} {\bibfnamefont
			{J.}~\bibnamefont {Myatt}},  \emph {et~al.},\ }\bibfield  {title} {\enquote
		{\bibinfo {title} {Direct-drive inertial confinement fusion: A review},}\
	}\href@noop {} {\bibfield  {journal} {\bibinfo  {journal} {Phys. Plasmas}\
		}\textbf {\bibinfo {volume} {22}},\ \bibinfo {pages} {110501} (\bibinfo
		{year} {2015})}\BibitemShut {NoStop}%
	\bibitem [{\citenamefont {Neutze}\ \emph {et~al.}(2000)\citenamefont {Neutze},
		\citenamefont {Wouts}, \citenamefont {van~der Spoel}, \citenamefont
		{Weckert},\ and\ \citenamefont {Hajdu}}]{Neutze00}%
	\BibitemOpen
	\bibfield  {author} {\bibinfo {author} {\bibfnamefont {R.}~\bibnamefont
			{Neutze}}, \bibinfo {author} {\bibfnamefont {R.}~\bibnamefont {Wouts}},
		\bibinfo {author} {\bibfnamefont {D.}~\bibnamefont {van~der Spoel}}, \bibinfo
		{author} {\bibfnamefont {E.}~\bibnamefont {Weckert}}, \ and\ \bibinfo
		{author} {\bibfnamefont {J.}~\bibnamefont {Hajdu}},\ }\bibfield  {title}
	{\enquote {\bibinfo {title} {Potential for biomolecular imaging with
				femtosecond x-ray pulses},}\ }\href@noop {} {\bibfield  {journal} {\bibinfo
			{journal} {Nature}\ }\textbf {\bibinfo {volume} {406}},\ \bibinfo {pages}
		{752--757} (\bibinfo {year} {2000})}\BibitemShut {NoStop}%
	\bibitem [{\citenamefont {Chapman}\ \emph {et~al.}(2011)\citenamefont
		{Chapman}, \citenamefont {Fromme}, \citenamefont {Barty}, \citenamefont
		{White}, \citenamefont {Kirian}, \citenamefont {Aquila}, \citenamefont
		{Hunter}, \citenamefont {Schulz}, \citenamefont {DePonte}, \citenamefont
		{Weierstall} \emph {et~al.}}]{Chapman11}%
	\BibitemOpen
	\bibfield  {author} {\bibinfo {author} {\bibfnamefont {H.~N.}\ \bibnamefont
			{Chapman}}, \bibinfo {author} {\bibfnamefont {P.}~\bibnamefont {Fromme}},
		\bibinfo {author} {\bibfnamefont {A.}~\bibnamefont {Barty}}, \bibinfo
		{author} {\bibfnamefont {T.~A.}\ \bibnamefont {White}}, \bibinfo {author}
		{\bibfnamefont {R.~A.}\ \bibnamefont {Kirian}}, \bibinfo {author}
		{\bibfnamefont {A.}~\bibnamefont {Aquila}}, \bibinfo {author} {\bibfnamefont
			{M.~S.}\ \bibnamefont {Hunter}}, \bibinfo {author} {\bibfnamefont
			{J.}~\bibnamefont {Schulz}}, \bibinfo {author} {\bibfnamefont {D.~P.}\
			\bibnamefont {DePonte}}, \bibinfo {author} {\bibfnamefont {U.}~\bibnamefont
			{Weierstall}},  \emph {et~al.},\ }\bibfield  {title} {\enquote {\bibinfo
			{title} {Femtosecond x-ray protein nanocrystallography},}\ }\href@noop {}
	{\bibfield  {journal} {\bibinfo  {journal} {Nature}\ }\textbf {\bibinfo
			{volume} {470}},\ \bibinfo {pages} {73--77} (\bibinfo {year}
		{2011})}\BibitemShut {NoStop}%
	\bibitem [{\citenamefont {Weber}\ \emph {et~al.}(2013)\citenamefont {Weber},
		\citenamefont {Riconda}, \citenamefont {Lancia}, \citenamefont {Marques},
		\citenamefont {Mourou},\ and\ \citenamefont {Fuchs}}]{Weber13}%
	\BibitemOpen
	\bibfield  {author} {\bibinfo {author} {\bibfnamefont {S.}~\bibnamefont
			{Weber}}, \bibinfo {author} {\bibfnamefont {C.}~\bibnamefont {Riconda}},
		\bibinfo {author} {\bibfnamefont {L.}~\bibnamefont {Lancia}}, \bibinfo
		{author} {\bibfnamefont {J.-R.}\ \bibnamefont {Marques}}, \bibinfo {author}
		{\bibfnamefont {G.}~\bibnamefont {Mourou}}, \ and\ \bibinfo {author}
		{\bibfnamefont {J.}~\bibnamefont {Fuchs}},\ }\bibfield  {title} {\enquote
		{\bibinfo {title} {Amplification of ultrashort laser pulses by {B}rillouin
				backscattering in plasmas},}\ }\href@noop {} {\bibfield  {journal} {\bibinfo
			{journal} {Phys. Rev. Lett.}\ }\textbf {\bibinfo {volume} {111}},\ \bibinfo
		{pages} {055004} (\bibinfo {year} {2013})}\BibitemShut {NoStop}%
	\bibitem [{\citenamefont {Hay}, \citenamefont {Valeo},\ and\ \citenamefont
		{Fisch}(2013)}]{Hay13}%
	\BibitemOpen
	\bibfield  {author} {\bibinfo {author} {\bibfnamefont {M.~J.}\ \bibnamefont
			{Hay}}, \bibinfo {author} {\bibfnamefont {E.~J.}\ \bibnamefont {Valeo}}, \
		and\ \bibinfo {author} {\bibfnamefont {N.~J.}\ \bibnamefont {Fisch}},\
	}\bibfield  {title} {\enquote {\bibinfo {title} {Geometrical optics of dense
				aerosols: Forming dense plasma slabs},}\ }\href {\doibase
		10.1103/PhysRevLett.111.188301} {\bibfield  {journal} {\bibinfo  {journal}
			{Phys. Rev. Lett.}\ }\textbf {\bibinfo {volume} {111}},\ \bibinfo {pages}
		{188301} (\bibinfo {year} {2013})}\BibitemShut {NoStop}%
	\bibitem [{\citenamefont {Malkin}, \citenamefont {Toroker},\ and\ \citenamefont
		{Fisch}(2014)}]{Malkin14}%
	\BibitemOpen
	\bibfield  {author} {\bibinfo {author} {\bibfnamefont {V.}~\bibnamefont
			{Malkin}}, \bibinfo {author} {\bibfnamefont {Z.}~\bibnamefont {Toroker}}, \
		and\ \bibinfo {author} {\bibfnamefont {N.}~\bibnamefont {Fisch}},\ }\bibfield
	{title} {\enquote {\bibinfo {title} {Exceeding the leading spike intensity
				and fluence limits in backward {R}aman amplifiers},}\ }\href@noop {}
	{\bibfield  {journal} {\bibinfo  {journal} {Phys. Rev. E}\ }\textbf {\bibinfo
			{volume} {90}},\ \bibinfo {pages} {063110} (\bibinfo {year}
		{2014})}\BibitemShut {NoStop}%
	\bibitem [{\citenamefont {Barth}\ \emph {et~al.}(2016)\citenamefont {Barth},
		\citenamefont {Toroker}, \citenamefont {Balakin},\ and\ \citenamefont
		{Fisch}}]{Barth16}%
	\BibitemOpen
	\bibfield  {author} {\bibinfo {author} {\bibfnamefont {I.}~\bibnamefont
			{Barth}}, \bibinfo {author} {\bibfnamefont {Z.}~\bibnamefont {Toroker}},
		\bibinfo {author} {\bibfnamefont {A.~A.}\ \bibnamefont {Balakin}}, \ and\
		\bibinfo {author} {\bibfnamefont {N.~J.}\ \bibnamefont {Fisch}},\ }\bibfield
	{title} {\enquote {\bibinfo {title} {Beyond nonlinear saturation of backward
				{R}aman amplifiers},}\ }\href@noop {} {\bibfield  {journal} {\bibinfo
			{journal} {Phys. Rev. E}\ }\textbf {\bibinfo {volume} {93}},\ \bibinfo
		{pages} {063210} (\bibinfo {year} {2016})}\BibitemShut {NoStop}%
	\bibitem [{\citenamefont {Jia}\ \emph {et~al.}(2017)\citenamefont {Jia},
		\citenamefont {Shi}, \citenamefont {Qin},\ and\ \citenamefont
		{Fisch}}]{Jia17}%
	\BibitemOpen
	\bibfield  {author} {\bibinfo {author} {\bibfnamefont {Q.}~\bibnamefont
			{Jia}}, \bibinfo {author} {\bibfnamefont {Y.}~\bibnamefont {Shi}}, \bibinfo
		{author} {\bibfnamefont {H.}~\bibnamefont {Qin}}, \ and\ \bibinfo {author}
		{\bibfnamefont {N.~J.}\ \bibnamefont {Fisch}},\ }\bibfield  {title} {\enquote
		{\bibinfo {title} {Kinetic simulations of laser parametric amplification in
				magnetized plasmas},}\ }\href@noop {} {\bibfield  {journal} {\bibinfo
			{journal} {Phys. Plasmas}\ }\textbf {\bibinfo {volume} {24}},\ \bibinfo
		{pages} {093103} (\bibinfo {year} {2017})}\BibitemShut {NoStop}%
	\bibitem [{\citenamefont {Karney}(1978)}]{Karney78}%
	\BibitemOpen
	\bibfield  {author} {\bibinfo {author} {\bibfnamefont {C.~F.}\ \bibnamefont
			{Karney}},\ }\bibfield  {title} {\enquote {\bibinfo {title} {Stochastic ion
				heating by a lower hybrid wave},}\ }\href@noop {} {\bibfield  {journal}
		{\bibinfo  {journal} {Phys. Fluids}\ }\textbf {\bibinfo {volume} {21}},\
		\bibinfo {pages} {1584--1599} (\bibinfo {year} {1978})}\BibitemShut {NoStop}%
	\bibitem [{\citenamefont {Karney}(1979)}]{Karney79}%
	\BibitemOpen
	\bibfield  {author} {\bibinfo {author} {\bibfnamefont {C.~F.}\ \bibnamefont
			{Karney}},\ }\bibfield  {title} {\enquote {\bibinfo {title} {Stochastic ion
				heating by a lower hybrid wave: Ii},}\ }\href@noop {} {\bibfield  {journal}
		{\bibinfo  {journal} {Phys. Fluids}\ }\textbf {\bibinfo {volume} {22}},\
		\bibinfo {pages} {2188--2209} (\bibinfo {year} {1979})}\BibitemShut {NoStop}%
	\bibitem [{\citenamefont {Sagdeev}\ and\ \citenamefont
		{Shapiro}(1973)}]{Sagdeev73}%
	\BibitemOpen
	\bibfield  {author} {\bibinfo {author} {\bibfnamefont {R.}~\bibnamefont
			{Sagdeev}}\ and\ \bibinfo {author} {\bibfnamefont {V.}~\bibnamefont
			{Shapiro}},\ }\bibfield  {title} {\enquote {\bibinfo {title} {Influence of
				transverse magnetic field on landau damping},}\ }\href@noop {} {\bibfield
		{journal} {\bibinfo  {journal} {JETP Lett.}\ }\textbf {\bibinfo {volume}
			{17}},\ \bibinfo {pages} {279--282} (\bibinfo {year} {1973})}\BibitemShut
	{NoStop}%
	\bibitem [{\citenamefont {Dawson}\ \emph {et~al.}(1983)\citenamefont {Dawson},
		\citenamefont {Decyk}, \citenamefont {Huff}, \citenamefont {Jechart},
		\citenamefont {Katsouleas}, \citenamefont {Leboeuf}, \citenamefont {Lembege},
		\citenamefont {Martinez}, \citenamefont {Ohsawa},\ and\ \citenamefont
		{Ratliff}}]{Dawson83}%
	\BibitemOpen
	\bibfield  {author} {\bibinfo {author} {\bibfnamefont {J.~M.}\ \bibnamefont
			{Dawson}}, \bibinfo {author} {\bibfnamefont {V.~K.}\ \bibnamefont {Decyk}},
		\bibinfo {author} {\bibfnamefont {R.~W.}\ \bibnamefont {Huff}}, \bibinfo
		{author} {\bibfnamefont {I.}~\bibnamefont {Jechart}}, \bibinfo {author}
		{\bibfnamefont {T.}~\bibnamefont {Katsouleas}}, \bibinfo {author}
		{\bibfnamefont {J.~N.}\ \bibnamefont {Leboeuf}}, \bibinfo {author}
		{\bibfnamefont {B.}~\bibnamefont {Lembege}}, \bibinfo {author} {\bibfnamefont
			{R.~M.}\ \bibnamefont {Martinez}}, \bibinfo {author} {\bibfnamefont
			{Y.}~\bibnamefont {Ohsawa}}, \ and\ \bibinfo {author} {\bibfnamefont {S.~T.}\
			\bibnamefont {Ratliff}},\ }\bibfield  {title} {\enquote {\bibinfo {title}
			{Damping of large-amplitude plasma waves propagating perpendicular to the
				magnetic field},}\ }\href {\doibase 10.1103/PhysRevLett.50.1455} {\bibfield
		{journal} {\bibinfo  {journal} {Phys. Rev. Lett.}\ }\textbf {\bibinfo
			{volume} {50}},\ \bibinfo {pages} {1455--1458} (\bibinfo {year}
		{1983})}\BibitemShut {NoStop}%
	\bibitem [{\citenamefont {Holkundkar}, \citenamefont {Brodin},\ and\
		\citenamefont {Marklund}(2011)}]{Holkundkar11}%
	\BibitemOpen
	\bibfield  {author} {\bibinfo {author} {\bibfnamefont {A.}~\bibnamefont
			{Holkundkar}}, \bibinfo {author} {\bibfnamefont {G.}~\bibnamefont {Brodin}},
		\ and\ \bibinfo {author} {\bibfnamefont {M.}~\bibnamefont {Marklund}},\
	}\bibfield  {title} {\enquote {\bibinfo {title} {Wakefield generation in
				magnetized plasmas},}\ }\href {\doibase 10.1103/PhysRevE.84.036409}
	{\bibfield  {journal} {\bibinfo  {journal} {Phys. Rev. E}\ }\textbf {\bibinfo
			{volume} {84}},\ \bibinfo {pages} {036409} (\bibinfo {year}
		{2011})}\BibitemShut {NoStop}%
	\bibitem [{\citenamefont {Jia}(2016)}]{Jia}%
	\BibitemOpen
	\bibfield  {author} {\bibinfo {author} {\bibfnamefont {Q.}~\bibnamefont
			{Jia}},\ }\href@noop {} {}\bibinfo {howpublished} {private communication}
	(\bibinfo {year} {2016})\BibitemShut {NoStop}%
\end{thebibliography}
%

\end{document}